\documentclass[openacc]{rstransa}%%%%where rstrans is the template name
\definecolor{darkgreen}{rgb}{0.05, 0.5, 0.06}

\usepackage{caption}
\usepackage{subcaption}
\titlehead{Research}

\begin{document}

%%%% Article title to be placed here
\title{The asymptotic state of decaying turbulence}

\author{%%%% Author details
Akash Rodhiya and Katepalli R. Sreenivasan}

% \orcid{\,KRS}{0000-0002-3943-6827}

%%%%%%%%% Insert author address here
\address{New York University, New York, NY, USA}

\citearticle{}

\history{}
%%%% Subject entries to be placed here %%%%
\subject{fluid mechanics, applied mathematics, field theory, differential equations, mathematical physics, statistical physics}

%%%% Keyword entries to be placed here %%%%

%%%% Insert corresponding author and its email address}
\corres{Katepalli R. Sreenivasan\\
\email{katepalli.sreenivasan@nyu.edu}}

%%%% Abstract text to be placed here %%%%%%%%%%%%
\begin{abstract}
The long-time evolution of decaying homogeneous turbulence is a fundamental building block of the theory and modeling of turbulence. We investigate the problem by using a comprehensive suite of direct numerical simulations. The simulations cover initial\break Taylor microscale Reynolds numbers $Re_\lambda$ from 30 to 145, with multiple independent realizations obtained at each $Re_{\lambda}$ to ensure statistical robustness. The energy spectrum is initialized with the Birkhoff--Saffman (BS) form (with $E(k) \sim k^2$ for small $k$) in one case, and the Loitsianskii--Kolmogorov--Batchelor (LKB) form (with $E(k) \sim k^4$ for small $k$) in another. Simulations are performed for unprecedented durations, of the order of 200,000 initial eddy-turnover times in some instances. For both BS and LKB, the turbulent kinetic energy $En$ shows, after an initial transient, unambiguous power-law decay, $En \sim t^{-n}$, with nearly constant decay exponents $n$, whose values are consistent with past theoretical results (and thus not universal). We compute various length scales, second-order structure functions and the spectral form at large wavenumbers; we note that an initially set $-5/3$ slope disappears quickly, while a perceptible $-1$ power region appears. In\break particular, we compare the present findings with predictions from the recent theory for decaying turbulence developed by Migdal (Migdal 2026 {\em Philos. Trans. R. Soc. A} \textbf{384}, 20250032. (\href{http://dx.doi.org/10.1098/rsta.2025.0032}{doi:10.1098/rsta.2025.0032})). The agreement for the BS case is excellent except for the large-wavenumber spectrum. A general discussion and assessment of results is provided in terms of the putative universality of energy decay. A main conclusion is that the energy decay is significantly influenced by `boundary effects', and that universality likely manifests only when those effects are removed. Alternatively, it\enlargethispage{10pt} may be more useful to discuss the universality of enstrophy decay.

This article is part of the theme issue `Frontiers of turbulence and statistical physics'.
\end{abstract}

%%%%%%%%%% Insert the texts which can accommodate on first page in the tag "fmtext" %%%%%

%\begin{fmtext}

%%%% Unnumbered equation

%\end{fmtext}

%%%%%%%%%%%%%%% End of first page %%%%%%%%%%%%%%%%%%%%%

\maketitle

\section{Introduction}
Canonical flows such as forced homogeneous and isotropic turbulence (HIT), wall-bounded turbulence, and thermal convection have provided a rich mosaic of insights into turbulence dynamics. Their nominal simplicity notwithstanding, these flows are actually quite complex, and their full theoretical understanding is not yet within reach. On the other hand, the particular case of \textit{decaying} HIT is more amenable to a fuller comprehension {because of the absence of the interaction between forcing and all the excited scales}---but it is still a non-trivial problem because it encompasses two primary ingredients of turbulence: nonlinear energy transfer across scales and turbulent dissipation. Advancing the quantitative understanding of decaying turbulence is the purpose of this paper.

Although decaying turbulence is one of the oldest problems in the field, complete clarity on its asymptotic state, presumed to be of the form $En \sim t^{-n}$, where $En$ is the turbulent kinetic energy, has remained elusive for reasons discussed in \cite{john-P-john}. Briefly, theoretical predictions have been made only for two particular states governing the `permanence of large-scales'. The first is the Birkhoff--Saffman (BS) regime \cite{Birkhoff,Saffman} associated with the conservation of linear momentum invariants, which prescribes a low-wavenumber spectrum $E(k) \sim k^2$, yielding $n=6/5$ (see, especially, the delightful and short paper by Saffman \cite{Saffman_1967}). The second is the Loitsianskii--Kolmogorov--Batchelor (LKB) regime \cite{Loitsiansky,kolmogorov1941,Batchelor1956}, associated with the conservation of angular momentum and characterized by a steeper low-wavenumber spectrum $E(k) \sim k^4$, for which $n=10/7$ (see \cite{comte-corrsin-1966} for an insightful derivation of this result). In both instances, certain crucial assumptions underlie the theoretical results.
Empirical works have usually not exercised wilful control on the initial spectrum, and, not surprisingly, yield results that often depart considerably from the theoretical predictions just mentioned. Experimental measurements in grid turbulence (see, e.g.,~\cite{comte-corrsin-1966,Sreenivasan_et_al_1980,MohamedLaRue1990,kang2003,sinhuber_et_al_2015, zhao_et_al_2023}) and direct numerical simulations in periodic boxes (see, e.g.,~\cite{Ishida2006, Yoffe2018, mkv_decay}) have reported decay exponents ranging from $n \approx 1.15$ to $1.45$. As reviewed by John \textit{et al.} \cite{john-P-john}, this unsatisfactory situation arises partly because the putative power-law fits often depend on the use of a virtual origin, or cover limited periods of time or close vicinity of the turbulence-generating grid, or include data that are influenced by details of the turbulence source. These inconsistencies also stem from the fact that the decay data are influenced by the finite computational or experimental domains. For another critical review of the experimental results, see ~\cite{ladislav}.

The present work explicitly addresses the limitations of past simulations highlighted by John \textit{et al.} \cite{john-P-john}. First, this is the only systematic effort to simulate the flow for unprecedented durations---exceeding 200,000 eddy turnover times in certain instances---ensuring the flow has the opportunity to reach a truly asymptotic state. Second, the initial conditions are strictly controlled at the spectral level, particularly regarding the low-wavenumber part. Third, the computational grid resolution is maintained high throughout the decay. We use a dynamic grid modification strategy but ensure that the resolution parameter $k_{\max}\eta$ {(where $k_{\max}$ is the highest wavenumber resolved and $\eta$ is the Kolmogorov lengthscale)} never falls below 3---double the standard requirement for faithful DNS---and often attains values exceeding 10 at late times, ensuring that the smallest scales of motion are always well resolved. With these robust datasets, we are now in a position to rigorously test the recent theoretical framework proposed by Migdal \cite{Migdal_theory}, which offers a parameter-free solution to the problem. While some questions remain, as discussed in \hyperref[sec:migdal_comp]{\S}\ref{sec:migdal_comp}, we show here that the agreement between the theory and simulations is remarkable for the BS case. For the LKB case, our results highlight a significant distinction between the universality of the spectral shape and the non-universality of the decay rate.

\hyperref[sec:simulations]{Section}~\ref{sec:simulations} presents a few details of simulations (including the grid modification method), while \hyperref[sec:results]{\S}\ref{sec:results} contains the principal results on the decay exponent, the evolution of
characteristic length scales, and the time evolution of the energy spectra. Finally, in \hyperref[sec:migdal_comp]{\S}\ref{sec:migdal_comp}, we compare our DNS findings with the
predictions of Migdal's theory, focusing on its predictions for decay exponents, length scales, spectral slopes {and enstrophy}. The theory is given in just enough detail to make intelligible comparisons, made in the same section, between theory and simulations. {Section 5 is a sensitivity analysis of various `bulk parameters' to changes in the low wavenumber truncation of the energy spectrum, primarily as a means to understand the `boundary effects' on energy decay.} The paper concludes with summary remarks in \hyperref[sec:conclusion]{\S}\ref{sec:conclusion}. By means of these systematic simulations, we particularly assess how the initial large-scale structure controls the decay.

% \newpage
\section{Simulations}
\label{sec:simulations}
\subsection{Initial conditions and simulation parameters}
Direct numerical simulations (DNS) were performed in a three-dimensional cubic domain with periodic boundary conditions. We employed a pseudospectral solver \cite{orszag1972, hussaini2007}, utilizing the phase-shift method for de-aliasing, which retains Fourier modes up to $k_{\max} \approx \sqrt{2}N/3$, where $N$ is the total number of wavenumbers in each direction \cite{Ishida2006}. Time integration was carried out using a semi-implicit predictor-corrector method.

In the previous DNS studies cited above, it has been customary to initialize the flow using an energy spectrum of the form $E(k) \sim (k/k_p)^q \exp[-(k/k_p)^2]$, where $q$ fixes the low-wavenumber scaling. John \textit{et al.} \cite{john-P-john} adopted a variant of this approach by modifying the low-wavenumber range of a fully developed forced turbulence spectrum. In the present work, we adopt the model spectrum of Pope \cite{Pope_2000} to provide more traditional control over the spectral shape. The velocity field was initialized as a random Gaussian field with the energy spectrum prescribed as \cite{rogallo1981numerical}
\begin{equation}
    E(k) = C\epsilon^{2/3}k^{-5/3}f_L(kL_e) \, f_{\eta}(k\eta),
    \label{eq:modal_spect}
\end{equation}
where $\epsilon$ is the energy dissipation rate, $k$ is the wavenumber, and $f_L$ and $f_{\eta}$ are non-dimensional shape functions for the energy-containing and dissipation ranges, respectively:
\begin{equation}
    f_L(kL_e) = \left(\frac{kL_e}{\left[(kL_e)^6 + c_L\right]^{1/6}}\right)^{5/3 + p_0},
    \label{eq:f_l}
\end{equation}\vskip-5pt
\begin{equation}
    f_{\eta}(k\eta) = \exp\left(-\beta \left(\left[(k\eta)^4 + c_{\eta}^4\right]^{1/4}-c_{\eta}\right)\right).
    \label{eq:f_eta}
\end{equation}
The exponent $p_0$ governs the spectral slope at low wavenumbers; $p_0=2$ for the BS spectrum and $p_0=4$ for the LKB spectrum. The parameter $L_e$ characterizes the large scales, determining the transition from $k^{p_0}$ to the inertial $k^{-5/3}$ range (typically $L_e \approx 2L$ to $3L$, where $L$ is the integral scale), while $\eta$ parametrizes the Kolmogorov dissipation scale; $c_L$, $c_\eta$ and $\beta$ are suitably assigned constants.

\begin{table}
\centering
\begin{tabular}{|l|l|l|l|l|l|l|}
\hline
Cases & $Re_{\lambda, t=0}$ & Grid & Runs & ${L_{t=0}}/{L_{box}}$ & \multicolumn{1}{c|}{\begin{tabular}[c]{@{}c@{}}Resolution$_{t=0}$\\ ($k_{max}\eta$)\end{tabular}} & \multicolumn{1}{c|}{\begin{tabular}[c]{@{}c@{}}Simulation time\\ ($t/T_{eddy,0}$)\end{tabular}}\\ \hline
1 & 30 & $512^3$ & 7 & 1.80$\%$ & 1.10 & $12418$ to $15098$ \\   \hline
2 & 45 & $1024^3$ & 7 & 1.49$\%$ & 1.20 & $36054$ to $41731$ \\  \hline
3 & 70 & $2048^3$ & 5 & 1.27$\%$ & 1.23 & $76695$ to $97074$ \\  \hline
4 & 93 & $2048^3$ & 10 & 1.21$\%$ & 0.85 & $70168$ to $90965$ \\  \hline
5 & 105 & $4096^3$ & 3 & 0.58$\%$ & 0.74 &  $184000$ to $264475$\\  \hline
6 & 145 & $4096^3$ & 5 & 1.11$\%$ & 0.86 & $96852$ to $238935$ \\  \hline
\end{tabular}
\caption{Simulation parameters for the BS spectra with $E(k) \sim k^2$ for small $k$.}
\label{tab:sims_k2}

\centering
\begin{tabular}{|l|l|l|l|l|l|l|}
\hline
Cases & $Re_{\lambda, t=0}$ & Grid & Runs  & ${L_{t=0}}/{L_{box}}$ & \multicolumn{1}{c|}{\begin{tabular}[c]{@{}c@{}}Resolution$_{t=0}$\\ ($k_{max}\eta$)\end{tabular}} & \multicolumn{1}{c|}{\begin{tabular}[c]{@{}c@{}}Simulation time\\ ($t/T_{eddy,0}$)\end{tabular}}\\ \hline
1 & 93 & $2048^3$ & 4 &  1.11\% & 0.85 & $70168$ to $90965$ \\ \hline
2 & 105 & $4096^3$ & 3 & 0.54\% & 0.74 & $163959$ to $185492$ \\ \hline
3 & 145 & $4096^3$ & 4 & 1.01\% & 0.87 & $105166$ to $138475$ \\ \hline
\end{tabular}
\caption{Simulation parameters for the LKB spectra with $E(k) \sim k^4$ for small $k$.}
\label{tab:sims_k4}

\end{table}
The complete set of simulation parameters is provided in \hyperref[tab:sims_k2]{tables}~\ref{tab:sims_k2} and~\ref{tab:sims_k4}, corresponding to the BS and LKB initializations, respectively. The tables list the initial Taylor-microscale Reynolds number, defined as $Re_{\lambda} = u'\lambda/\nu$ (where $u' = \sqrt{\langle u_iu_i\rangle/3}$ and $\lambda = \sqrt{u'^2/\langle (\partial u/\partial x)^2 \rangle}$), which ranges from 30 to 145. The integral scale $L$ is defined as $(3\pi/4)(\int_0^{\infty}k^{-1}E(k)dk)/(\int_0^{\infty}E(k)dk)$.
To mitigate finite-domain effects, we ensured that the initial integral scale $L$ remained small relative to the box size ($L_{\mathrm{box}}=2\pi$). For the majority of runs, the energetic length scale was fixed at $L_e = L_{\mathrm{box}}/35$, resulting in $L \approx 1\%$ of $L_{\mathrm{box}}$. However, for the high Reynolds number cases (case 5 in \hyperref[tab:sims_k2]{table}~\ref{tab:sims_k2} and case 2 in \hyperref[tab:sims_k4]{table}~\ref{tab:sims_k4}), we further lowered $L_e$ to $L_{\mathrm{box}}/70$. This modification yielded initial integral scales as small as $0.58\%$ and $0.54\%$ of the domain size, respectively.
With increasing $Re_{\lambda}$, the computational mesh size increases to resolve the smallest scales. The initial grid resolution $k_{\max}\eta$ is approximately unity (see \hyperref[tab:sims_k2]{tables}~\ref{tab:sims_k2} and~\ref{tab:sims_k4}). As the turbulence decays, $\eta$ grows, naturally increasing the effective resolution; consequently, $k_{\max}\eta$ continually exceeds unity, ensuring that the flow remains well resolved. The simulation durations are unprecedented, extending up to $2.0 \times 10^5$ initial eddy turnover times ($T_{\mathrm{eddy},0} = L_{t=0}/u'_{t=0}$) for the highest Reynolds number cases. We performed multiple simulation runs to generate an adequate ensemble for statistical convergence; the number of realizations is listed in the `Runs' column, with each realization differing by the random seed used to generate the initial velocity field.

\begin{figure}
    % \centering
    \begin{subfigure}[b]{0.5\textwidth} % Adjust width as needed
        \centering
        \includegraphics[width=\textwidth]{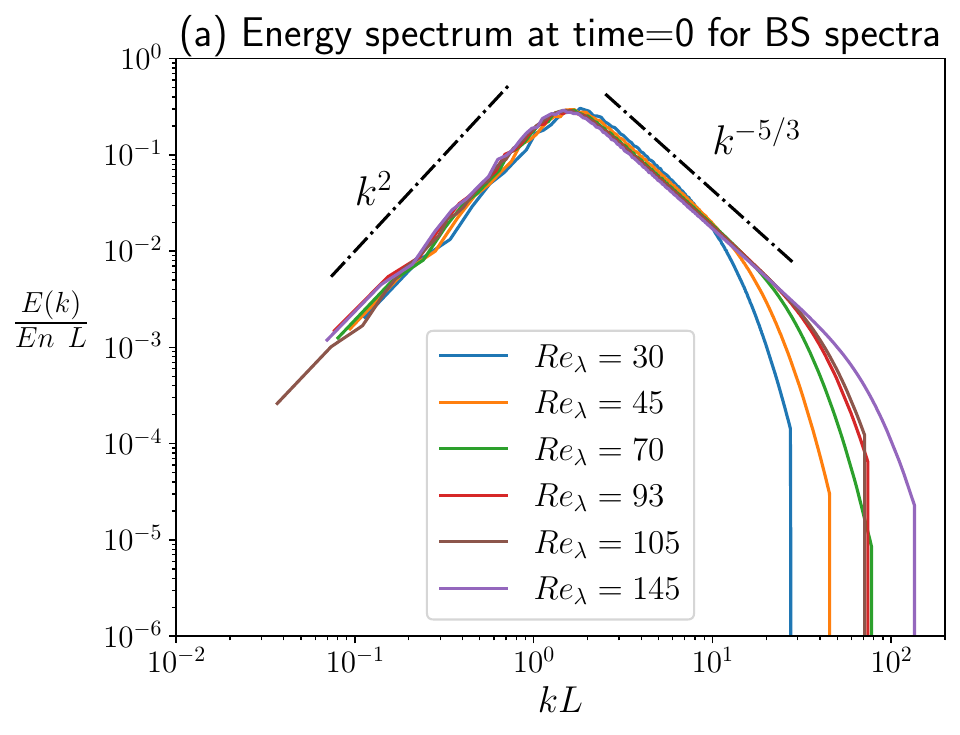} 
        % \put(-150,125){\normalsize{(a)}}
        % \caption{}
        % \label{fig:decayExp_k2_a}
    \end{subfigure}
    \hfill % Adds horizontal space between subfigures
    \begin{subfigure}[b]{0.5\textwidth} % Adjust width as needed
        \centering
        \includegraphics[width=\linewidth]{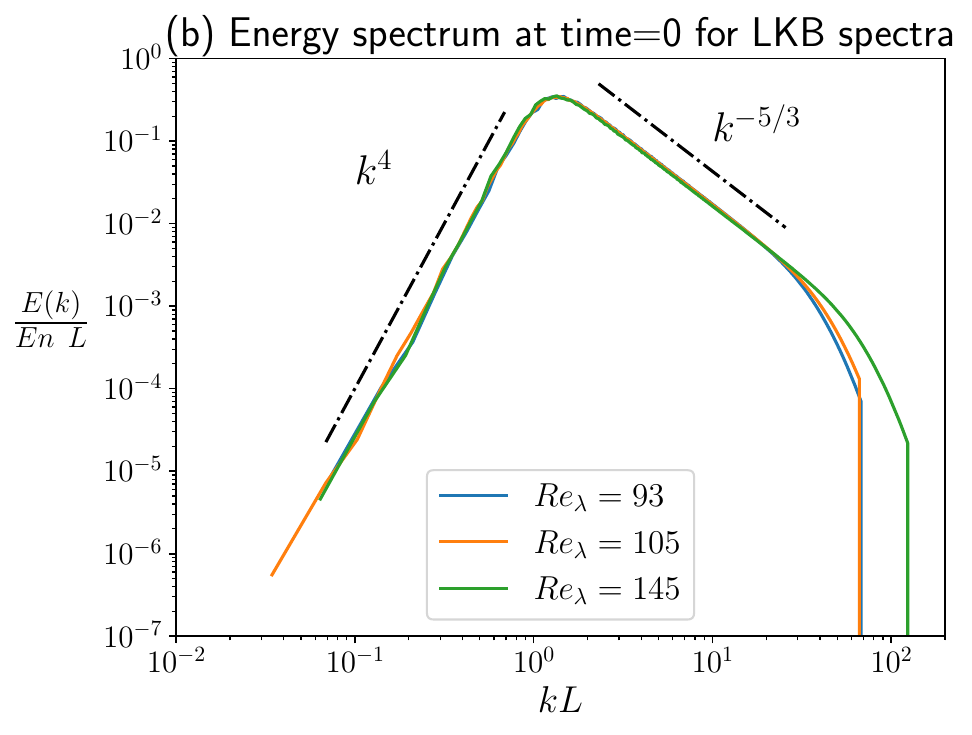}
        % \put(-150,125){\normalsize{(b)}}
        % \label{fig:decayExp}
    \end{subfigure}
    \caption{Initial energy spectra ($t=0$) used for the simulations. The spectra are normalized by the initial integral length $L$ and initial total energy $En(t=0)$. The wavenumber $k$ is normalized by $L$. (a) BS spectra with the prescribed $E(k) \sim k^2$ scaling at low wavenumbers. (b) LKB spectra with the prescribed $E(k) \sim k^4$ scaling at low wavenumbers. Different colored curves correspond to different initial Taylor-microscale Reynolds numbers ($Re_\lambda$) for each set of simulations. The dot-dashed lines indicate the theoretical slopes for the low-wavenumber range and the $k^{-5/3}$ inertial range.}
    \label{fig:init_spect}
\end{figure}

The initial energy spectra ($t=0$) for the BS (\hyperref[tab:sims_k2]{table}~\ref{tab:sims_k2}) and LKB (\hyperref[tab:sims_k4]{table}~\ref{tab:sims_k4}) simulations are plotted in \hyperref[fig:init_spect]{figure}~\ref{fig:init_spect}. Panel (a) shows the BS spectra, which follow a prescribed $k^2$ power law at low wavenumbers, while panel (b) shows the LKB spectra, which follow $k^4$. In both panels, the spectra are normalized by the initial total energy $En(t=0)$ and initial integral length $L$, with the wavenumber normalized by $L$. The dot-dashed reference lines confirm the prescribed low-wavenumber slopes ($k^2$ or $k^4$) and the $k^{-5/3}$ inertial range. For low wavenumbers, the spectra exhibit fluctuations because the number of discrete Fourier modes available in spherical shells at low $k$ is small. At higher wavenumbers, the mode density increases significantly and results in a smoother spectral profile. 

\subsection{Computational grid modification}
\begin{figure}[htbp]
    \centering
    % Left Column: The Table
    \begin{minipage}[c]{0.40\textwidth}
        \centering
        \begin{tabular}{|c|l|}
\hline
\begin{tabular}[c]{@{}c@{}}Threshold\\ ($k_{max}\eta$)\end{tabular} & \multicolumn{1}{c|}{\begin{tabular}[c]{@{}c@{}}Fractional \\ energy loss\end{tabular}} \\ \hline
1.8 & $8.0 \times 10^{-4}$    \\ \hline
2.0 & $3.0 \times 10^{-4}$    \\ \hline
2.2 & $2.5 \times 10^{-4}$    \\ \hline
3.0 & $2.3 \times 10^{-5}$    \\ \hline
5.0 & $6.3 \times 10^{-8}$    \\ \hline
6.0 & $5.0 \times 10^{-9}$    \\ \hline
\end{tabular}
        % \captionof allows a table caption inside a figure environment
        \captionof{table}{Quantification of fractional energy loss incurred due to spectral truncation at different regridding thresholds ($k_{max}\eta$) for $Re_{\lambda} = 93$.}
        \label{tab:gridModThreshold}
    \end{minipage}
    \hfill % Adds space between the columns
    % Right Column: The Figure
    \begin{minipage}[c]{0.48\textwidth}
        \centering
        % Replace with your actual file path
        \includegraphics[width=\linewidth]{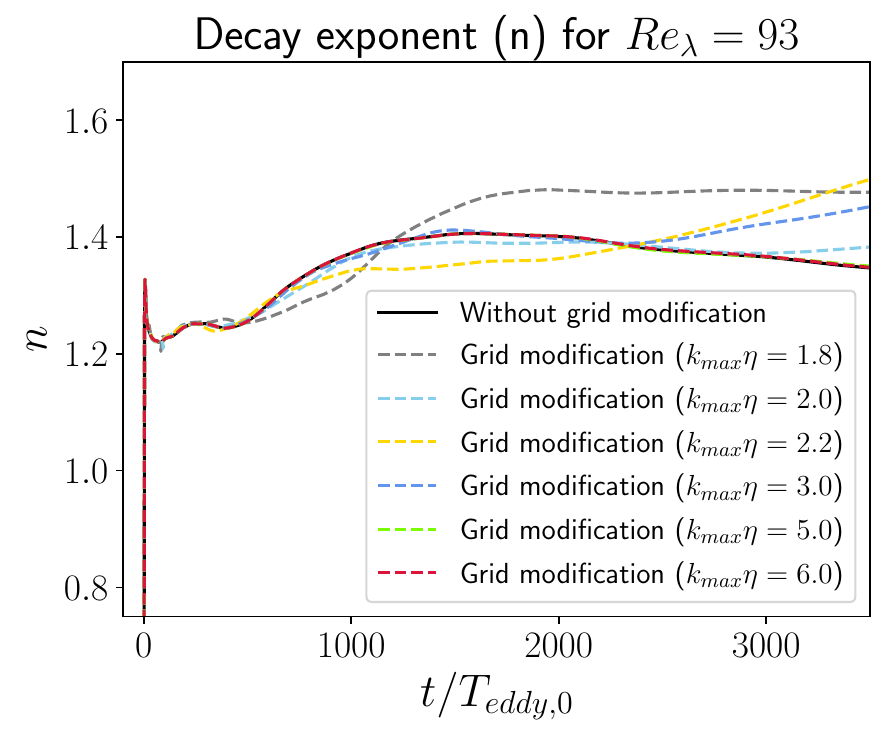} 
        \caption{Validation of the grid modification method for the $Re_\lambda = 93$ simulation. The plot compares the time evolution of the decay exponent $n(t)$ for the standard simulation (solid black line) against simulations using grid modification with various thresholds (dashed lines). The curves for lower thresholds depart measurably, whereas the $k_{max}\eta = 6.0$ case (red dashed line) overlaps perfectly with the unmodified simulation, demonstrating no statistical modification.}
        \label{fig:GM_comp}
    \end{minipage}
\end{figure}

Our key objective is to understand the long-time power laws (assuming that they exist). However, computational cost increases significantly for longer simulation times. While the time step $\Delta t$ can be increased as the energy decays owing to the loosening CFL constraint, it provides only a partial relief. In previous work, the simulation time was on the order of $1000$ eddy turnover times \cite{john-P-john, mkv_decay}; this range is substantially smaller than in the present work, and the status of the power laws was often unclear. We extend the simulations for longer times using the grid modification method described below.
As kinetic energy decays, the Kolmogorov length scale $\eta$ increases, rendering the initial discretization to become progressively over-resolved. To optimize computational efficiency, we implement a dynamic regridding scheme. We halve the grid resolution when the resolution parameter exceeds a specific threshold, thereby reducing the maximum resolved wavenumber $k_{\max}$ by a factor of two. This process is repeated until a grid size of $512^3$ is reached (this being the smallest grid size used), allowing us to extend some simulations ($Re_{\lambda} = 93$, $105$, and $145$) for extended durations beyond 200,000 the initial eddy turnover times. Each regridding event reduces the degrees of freedom by a factor of 8 and typically allows for a doubled time step, yielding a theoretical speedup of roughly 16 times per step.
However, regridding inherently involves the truncation of high-wavenumber Fourier modes (even though they are much finer than the Kolmogorov wavenumber). To ensure that this truncation does not alter the physical decay dynamics, the threshold for grid reduction must be chosen carefully. We performed a sensitivity analysis using a $2048^3$ simulation at $Re_{\lambda} = 93$, testing thresholds from $k_{\max}\eta = 1.8$ to $6.0$. \hyperref[tab:gridModThreshold]{Table}~\ref{tab:gridModThreshold} quantifies the fractional energy loss incurred during the truncation step for each threshold. For low thresholds (e.g. $k_{\max}\eta = 1.8$), the instantaneous energy loss is on the order of $8.0 \times 10^{-4}$, which shows a small fraction of physically active scales are being discarded. For $k_{\max}\eta = 6.0$, the energy loss drops to $5.0 \times 10^{-9}$, showing that the truncated modes contain even less energy.
\hyperref[fig:GM_comp]{Figure}~\ref{fig:GM_comp} illustrates the time evolution of the decay exponent $n$ for these various thresholds. The curves for $k_{\max}\eta < 6.0$ depart for long times from the reference simulation (no grid modification), showing that aggressive regridding artificially alters the decay rate. Conversely, the results for $k_{\max}\eta = 6.0$ are indistinguishable from the unmodified DNS. We thus fixed the regridding threshold at $k_{\max}\eta = 6$, which resets the new resolution to $k_{\max}\eta \approx 3$, ensuring that the simulations maintain high spectral fidelity throughout the decay process.

\section{Basic results}
\label{sec:results}
The BS and the LKB spectra correspond to distinct invariants derived from the K\'{a}rm\'{a}n--Howarth equations. The BS invariant corresponds to the conservation of linear momentum and the\break Loitsianskii invariant to the conservation of angular momentum---leading to different theoretical predictions for the decay rate and the growth of the large scale.

\subsection{Power laws for energy decay and length scale growth}
\label{subsec:power_laws}
\begin{figure}
    % \centering
    \begin{subfigure}[b]{0.5\textwidth} % Adjust width as needed
        \centering
        \includegraphics[width=\textwidth]{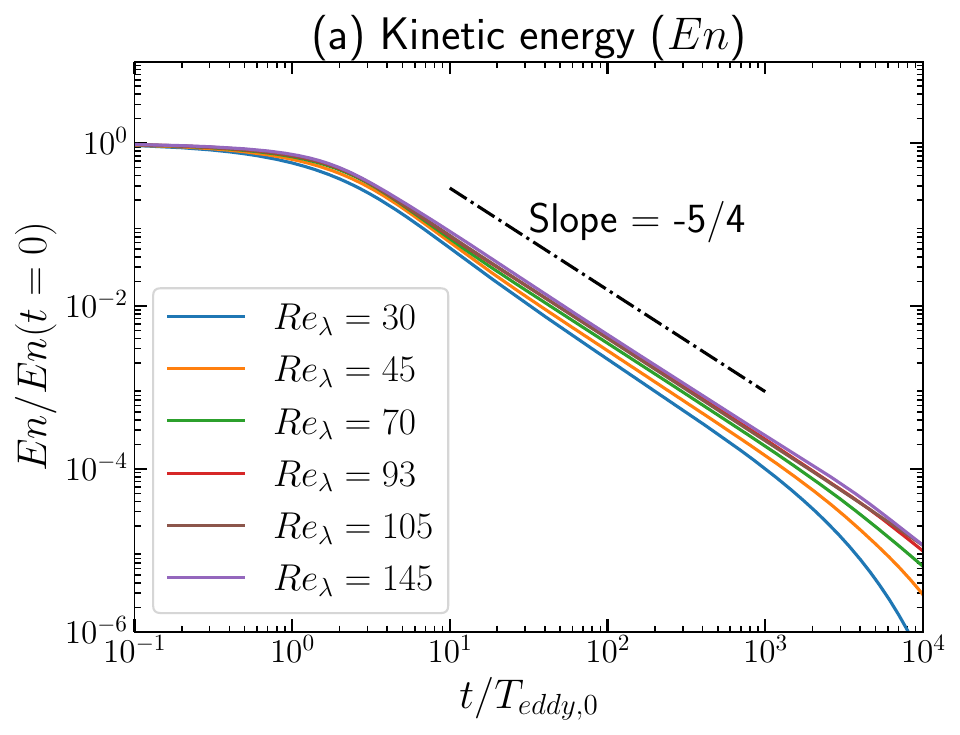} 
        % \put(-160,125){\normalsize{(a)}}
        % \caption{}
        % \label{fig:decayExp_k2_a}
    \end{subfigure}
    \hfill % Adds horizontal space between subfigures
    \begin{subfigure}[b]{0.5\textwidth} % Adjust width as needed
        \centering
        \includegraphics[width=\linewidth]{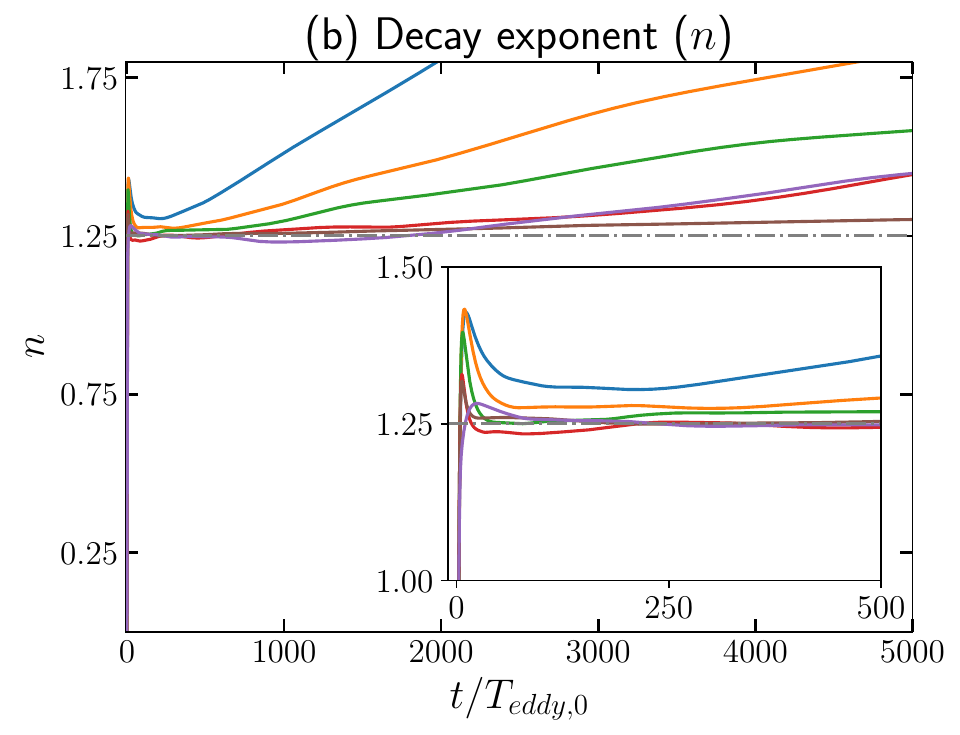}
        % \put(-160,125){\normalsize{(b)}}
        % \label{fig:decayExp}
    \end{subfigure}
    \caption{Time evolution of (a) total kinetic energy $En(t)$ and (b) its local decay exponent $n(t)$ for simulations with an initial BS spectrum ($E(k) \sim k^2$ for small $k$). The different-coloured curves in both panels correspond to simulations with varying initial Taylor Reynolds numbers ($Re_\lambda$), as indicated in the legends. The time axis is normalized by the initial large-eddy turnover time, $T_{eddy,0}$. (a) $En$ is normalized by its initial value $En(t=0)$. The dot-dashed reference line shows Migdal's theoretical power-law decay slope of $-5/4$ (to be described later). High Reynolds number data asymptote to that slope. (b) The local kinetic energy decay exponent, defined in Eq.~(\ref{eq:decay_exp}). The horizontal dot-dashed line marks $n = 5/4$ (which, to anticipate the results in Sec.\ 4, corresponds to Migdal's theory). The inset provides a magnified view of the early-time evolution, showing the approach towards 5/4. The index $n$ for the lowest Reynolds number slope does not touch 5/4 but it does so at the higher Reynolds numbers, with the range increasing correspondingly, as one may have expected.  }
    \label{fig:decayExp_k2}
\end{figure}

A basic question is whether the decaying turbulence follows a credible power law and, if so, for how long. For all cases, turbulence is initialized as a random Gaussian field prescribed with a given total kinetic energy and energy spectrum (see equations~\ref{eq:modal_spect}--\ref{eq:f_eta}). Since there is no forcing, the kinetic energy decays monotonically. \hyperref[fig:decayExp_k2]{Figure}~\ref{fig:decayExp_k2} shows the time evolution of the total kinetic energy (panel a) and the local decay exponent (panel b) for the simulations initialized with the BS spectra (see \hyperref[tab:sims_k2]{table} \ref{tab:sims_k2}). Each curve represents the ensemble average of all realizations for a given Reynolds number. The decay exponent is computed as
\begin{align}
    n = -\frac{d \log{En}}{d \log{t}} = -(1/En)\frac{dEn/dt}{1/t} = \frac{t\epsilon}{En},
    \label{eq:decay_exp}
\end{align}
where $\epsilon = \int_0^{k_{\max}} 2\nu k^2 E(k) dk$ is the energy dissipation rate.
In \hyperref[fig:decayExp_k2]{figure}~\ref{fig:decayExp_k2}, the kinetic energy is normalized by its initial value and plotted on logarithmic axes, while the decay exponent is plotted on a linear scale. Time is normalized by the initial large-eddy turnover time ($T_{\mathrm{eddy},0}$). \hyperref[fig:decayExp_k2]{Figure}~\ref{fig:decayExp_k2}\hyperref[fig:decayExp_k2]{a} shows that, after an initial transient of some 3 eddy turnover times, the kinetic energy for all cases decays with a nearly constant exponent, converging towards the slope indicated by the dot-dashed reference line.
\hyperref[fig:decayExp_k2]{Figure}~\ref{fig:decayExp_k2}\hyperref[fig:decayExp_k2]{b} shows the local slope computed via equation~\ref{eq:decay_exp}, while the inset is a magnified view of the early-time evolution ($0$--$500$ turnover times). The decay exponent $n$ maintains a distinct plateau for a significant duration. As seen in the inset, both the value of the exponent and the duration of this plateau depend on the Reynolds number. For $Re_{\lambda}=30$, we observe $n \approx 1.31$; this value is $1.28$, $1.26$ and $1.26$ for $Re_{\lambda}=45$, $70$ and $93$, respectively. For the highest two Reynolds numbers ($Re_\lambda = 105$ and $145$), the exponent settles down to $n \approx 1.25$.
The duration for which the decay exponent remains constant generally increases with the Reynolds number. The region of constant $n$ is followed by an increase for all cases.  

\begin{figure}
    % \centering
    \begin{subfigure}[b]{0.5\textwidth} % Adjust width as needed
        \centering
        \includegraphics[width=\linewidth]{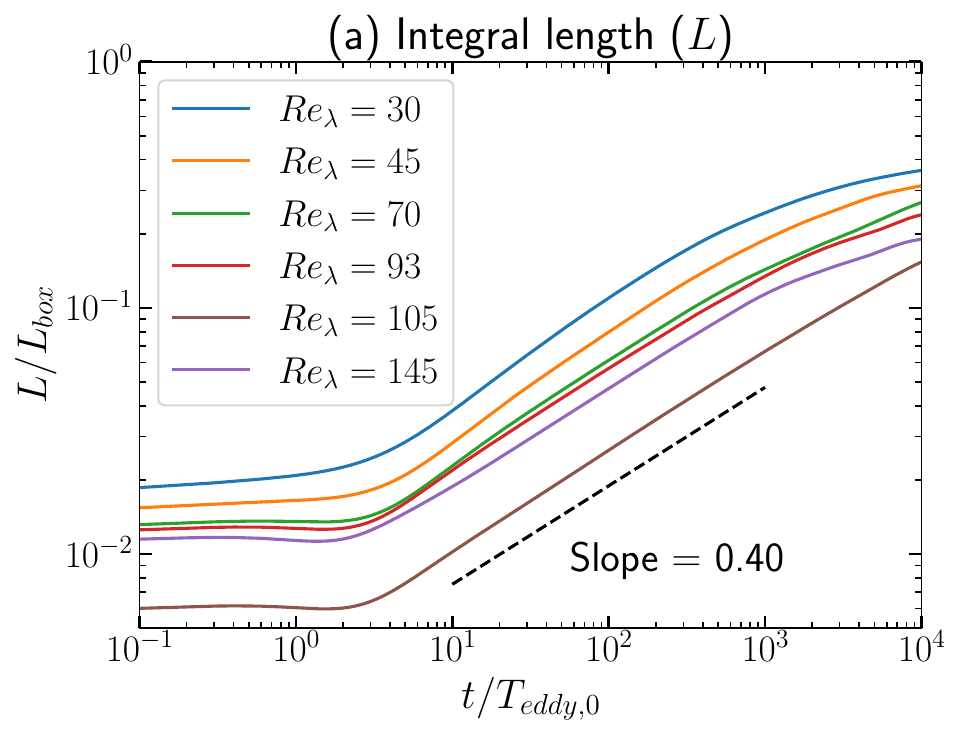}
        % \put(-20,25){\normalsize{(a)}}
        % \caption{}
        % \label{fig:IntLen}
    \end{subfigure}
    \hfill % Adds horizontal space between subfigures
    \begin{subfigure}[b]{0.5\textwidth} % Adjust width as needed
        \centering
        \includegraphics[width=\linewidth]{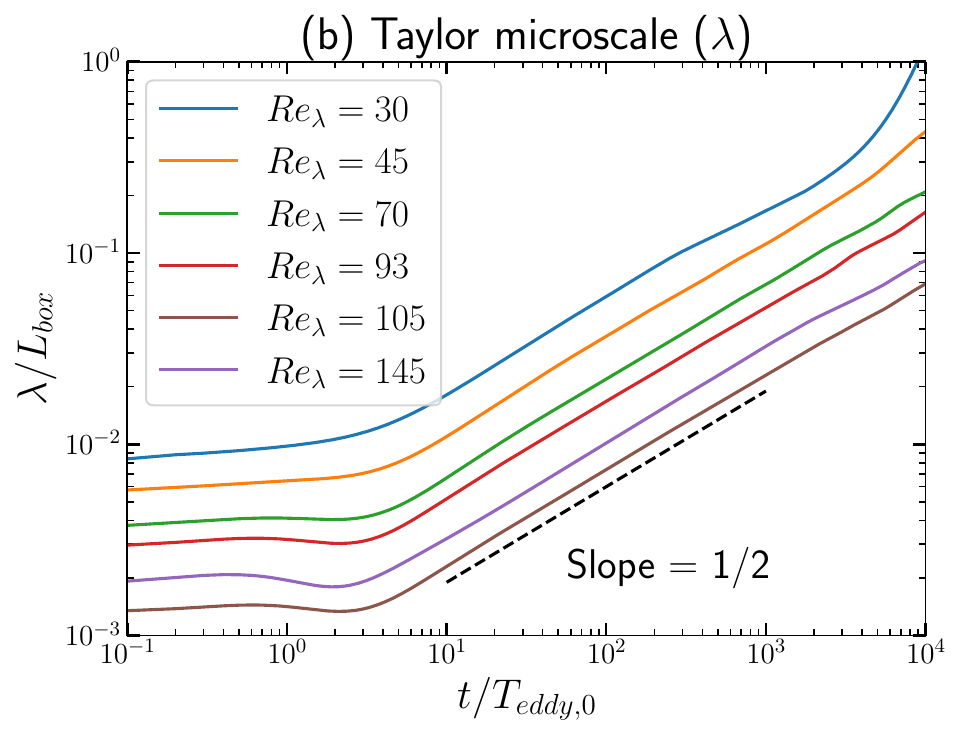}
        % \put(-20,25){\normalsize{(b)}}
        % \caption{}
        % \label{fig:lambda}
    \end{subfigure}
    \begin{subfigure}[b]{0.5\textwidth} % Adjust width as needed
        \centering
        \includegraphics[width=\linewidth]{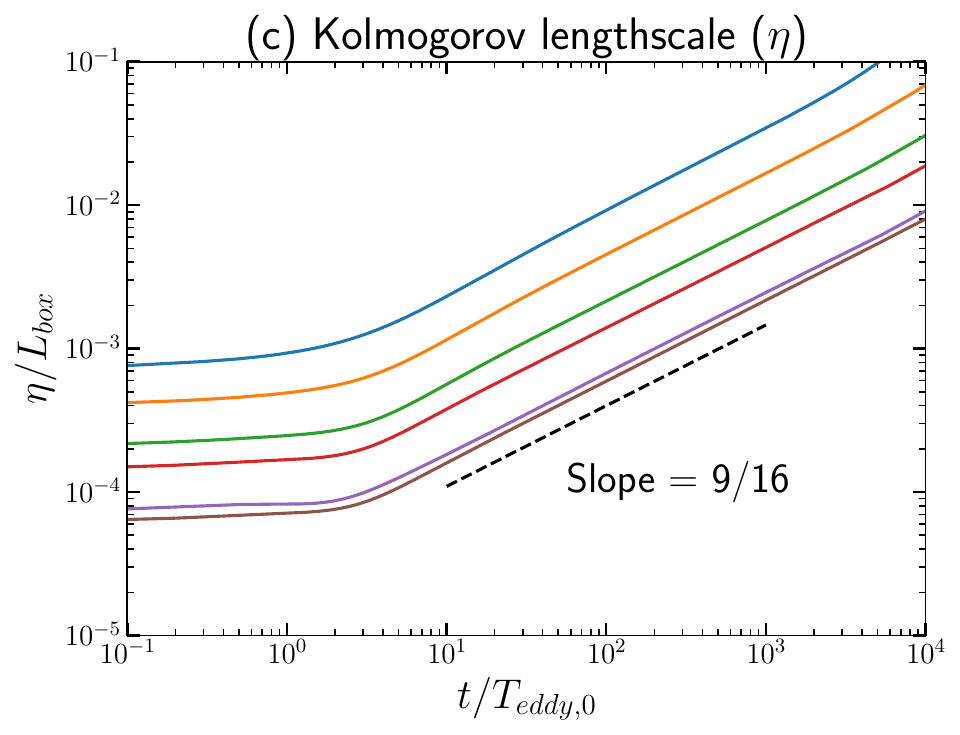}
        % \put(-20,25){\normalsize{(c)}}
        % \caption{}
        % \label{fig:kolmogorov_len}
    \end{subfigure}
    \hfill % Adds horizontal space between subfigures
    \begin{subfigure}[b]{0.5\textwidth} % Adjust width as needed
        \centering
        \includegraphics[width=\linewidth]{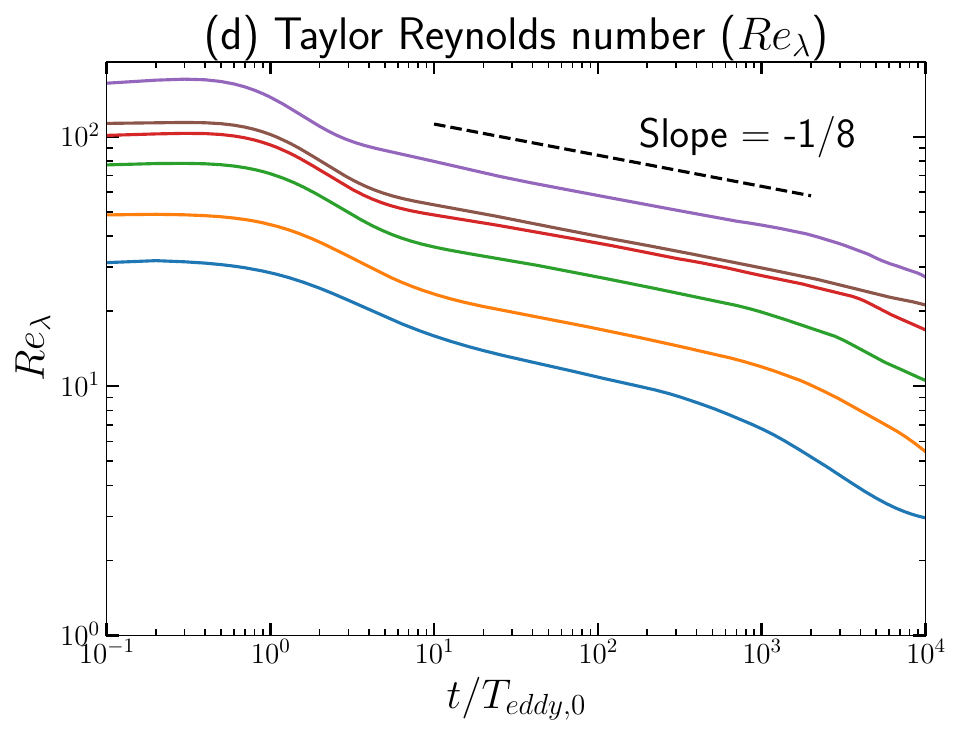}
        % \caption{}
        % \put(-20,25){\normalsize{(d)}}
        % \label{fig:re_lam}
    \end{subfigure}
    \caption{Time evolution of characteristic length scales and the Taylor Reynolds number for simulations with an initial BS spectrum ($E(k) \sim k^2$ for small $k$). All panels show results from simulations with the same initial $Re_\lambda$ (see legends), with time normalized by the initial large-eddy turnover time, $T_{eddy,0}$. The subplots show: (a) the integral length scale $L$, (b) the Taylor microscale $\lambda$, and (c) the Kolmogorov length scale $\eta$, each normalized by the edge length of the domain box ($L_{box}=2\pi$). (d) The unnormalized Taylor Reynolds number $Re_\lambda$. The dashed reference lines indicate the theoretical power-law behaviors: $L \sim t^{2/5}$ in (a), $\lambda \sim t^{1/2}$ in (b), $\eta \sim t^{9/16}$ in (c), and $Re_\lambda \sim t^{-1/8}$ in (d).}
    \label{fig:power_laws_k2}
\end{figure}

In addition to the energy decay, distinct power laws are observed in the time evolution of the characteristic length scales and the Reynolds number. \hyperref[fig:power_laws_k2]{Figure}~\ref{fig:power_laws_k2} presents the time evolution of (a) the integral length scale $L$, (b) the Taylor microscale $\lambda$, (c) the Kolmogorov length scale $\eta$ and (d) the microscale Reynolds number $Re_{\lambda}$, for the simulations initialized with the BS spectra (\hyperref[tab:sims_k2]{table}~\ref{tab:sims_k2}). All length scales are normalized by the box size ($L_{\mathrm{box}} = 2\pi$), and time is normalized by the initial large-eddy turnover time.
As shown in \hyperref[fig:power_laws_k2]{figure}~\ref{fig:power_laws_k2}\hyperref[fig:power_laws_k2]{a}, the normalized integral length scale scales as $L \sim t^{2/5}$, consistent with previous studies~\cite{john-P-john,mkv_decay}. Although theoretical frameworks often link this scaling to a decay exponent of $n=1.2$ (assuming $\epsilon \sim u'^3/L$)~\cite{john-P-john}, our simulations seem to favour a slightly higher decay exponent of $n \approx 1.25$.
The Taylor microscale (\hyperref[fig:power_laws_k2]{figure}~\ref{fig:power_laws_k2}\hyperref[fig:power_laws_k2]{b}) follows the power law $\lambda \sim t^{1/2}$, which is consistent with the definition $\lambda^2 \sim u'^2/\epsilon$ regardless of the specific exponent for the energy decay. Similarly, the Kolmogorov scale evolves as $\eta \sim t^{9/16}$ (\hyperref[fig:power_laws_k2]{figure}~\ref{fig:power_laws_k2}\hyperref[fig:power_laws_k2]{c}), consistent with the relation $\eta = (\nu^3/\epsilon)^{1/4}$ and the observed dissipation decay $\epsilon \sim t^{-9/4}$. Finally, the Reynolds number (\hyperref[fig:power_laws_k2]{figure}~\ref{fig:power_laws_k2}\hyperref[fig:power_laws_k2]{d}) follows $Re_{\lambda} \sim t^{-1/8}$, satisfying the relation $Re_\lambda = u' \lambda / \nu$.

The Reynolds number decays only slowly so that, if it is high enough to start with, it will remain {`adequately high' for the duration of decay (i.e. with no sight of the `final period of decay').} The main change that occurs during decay, apart from the energy itself, is the ratio of the integral length to the domain size, $L/L_{\mathrm{box}}$. As $L$ grows to approximately $10\%$--$15\%$ of the box size, finite-size effects emerge, resulting in deviations from power laws \cite{mkv_decay, thornber2016, Touil2002}. The small-scale $\eta$ remains unaffected because of its separation from the large scales.
The power laws persist longer for higher Reynolds numbers. This is because the initial integral length scale $L_{t=0}$ is smaller for larger $Re_\lambda$ ($1.1\%$ of $L_{\mathrm{box}}$ at $Re_\lambda=145$ compared to $1.8\%$ at $Re_\lambda=30$), providing more time for its growth before confinement effects set in. The $Re_\lambda=105$ case (case 5) is an outlier that exhibits an exceptionally long power-law regime. As detailed in \hyperref[sec:simulations]{\S}\ref{sec:simulations}, this case was initialized with an extended $k^2$ scaling range at low wavenumbers, resulting in a significantly smaller initial integral length ($0.58\%$ of $L_{\mathrm{box}}$) compared to typical cases. This supports the claim that a lower initial integral length (in relation to the box size) is the key factor in prolonging self-similar decay.

\begin{figure}
    % \centering
    \begin{subfigure}[b]{0.5\textwidth} % Adjust width as needed
        \centering
        \includegraphics[width=\linewidth]{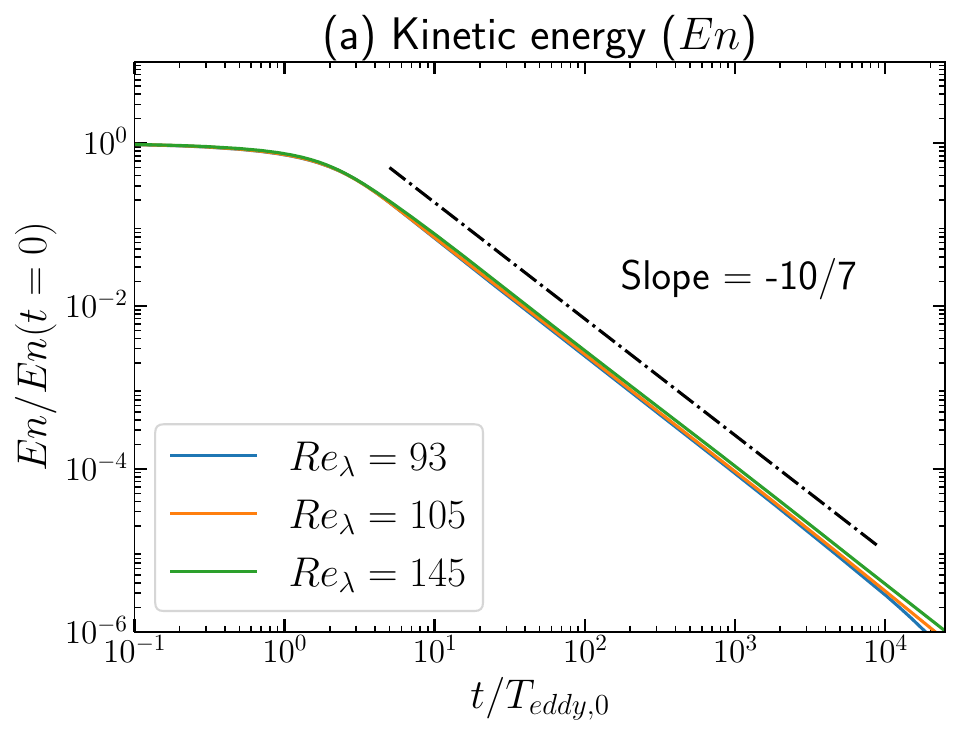}
        % \caption{}
        % \label{fig:totE_k4}
    \end{subfigure}
    \hfill % Adds horizontal space between subfigures
    \begin{subfigure}[b]{0.5\textwidth} % Adjust width as needed
        \centering
        \includegraphics[width=\linewidth]{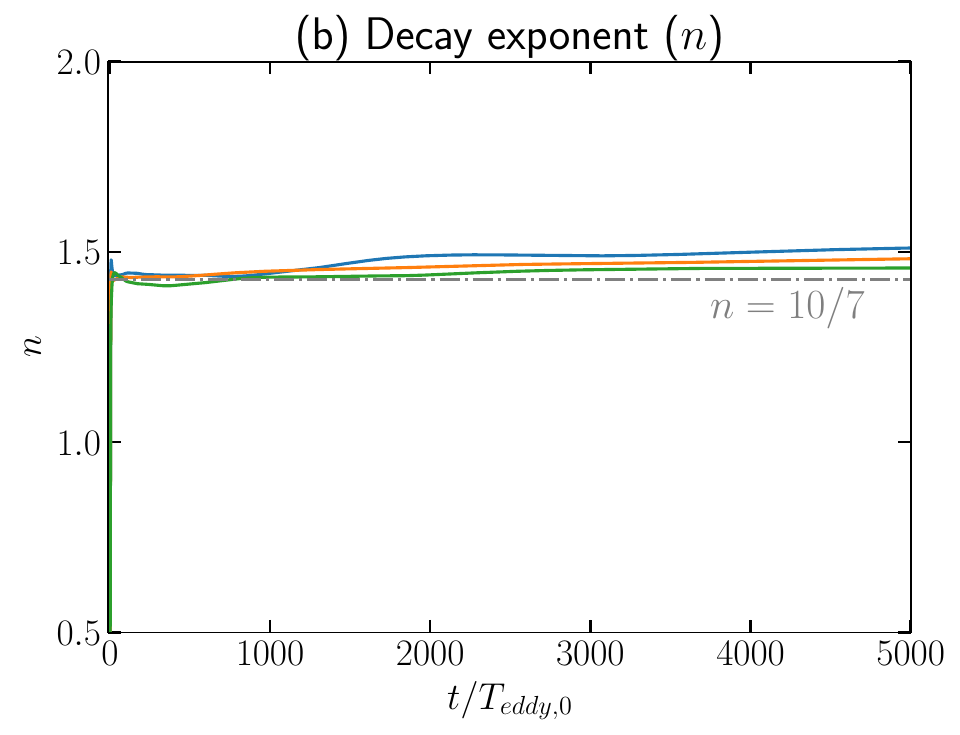}
        % \caption{}
        % \label{fig:decayExpK4}
    \end{subfigure}
    \caption{Time evolution of (a) normalized total kinetic energy $En(t)$ and (b) its local decay exponent $n(t)$ for simulations with an initial LKB spectrum ($E(k) \sim k^4$ for small $k$). Both panels show results for three different initial Taylor Reynolds numbers ($Re_\lambda = 93$, $105$ and $145$), with the time axis normalized by the initial large-eddy turnover time, $T_{eddy,0}$. (a) Total kinetic energy $En$, normalized by its initial value $En(t=0)$. The dot-dashed reference line indicates the theoretical power-law decay slope of $-10/7$. (b) The local kinetic energy decay exponent, defined in Eq.~(\ref{eq:decay_exp}). The horizontal dot-dashed line represents the Kolmogorov value $n = 10/7$.}
    \label{fig:decayExpK4}
\end{figure}

%LKB initialisation results
The energy decay dynamics are highly sensitive to the initial energy spectrum. (Other properties of initial turbulence may also matter for the decay but since we are interested in second-order quantities, we may regard the higher level information to be of secondary importance.) The LKB initialization evolves with a distinctly different decay exponent and scaling laws.
\hyperref[fig:decayExpK4]{Figure}~\ref{fig:decayExpK4} presents the time evolution of the kinetic energy (panel a) and the local decay exponent (panel b) for the LKB spectra initialized simulations. The kinetic energy and time are normalized by their respective initial values and eddy turnover time. Each curve represents the ensemble average for the chosen Reynolds number.

The LKB spectra simulations exhibit a decay exponent of $n \approx 10/7$, just as Kolmogorov had calculated on the basis of the Loitsianskii invariant. The dot-dashed reference line, with a slope of $-10/7$, shows excellent agreement with the simulation data in panel (a). As shown in panel (b), the local decay exponent $n$ remains close to Kolmogorov's value of $10/7$ for all cases. For $Re_{\lambda} = 93$, the exponent begins to depart from $10/7$ after approximately $1000$ eddy turnover times, whereas for the higher Reynolds numbers ($Re_{\lambda} = 105$ and $145$), the plateau is remarkably robust, persisting beyond $2000$ turnover times.
A notable difference compared to the BS spectra simulations is the persistence of the power-law regime. Although the BS cases showed a relatively short plateau at $  n\approx 1.25$ before drifting upwards, the LKB spectra simulations maintain a stable exponent for a significantly longer duration. As with the BS results, however, the region of constant $n$ breaks down eventually.

\begin{figure}
    \centering
    \includegraphics[width=0.5\linewidth]{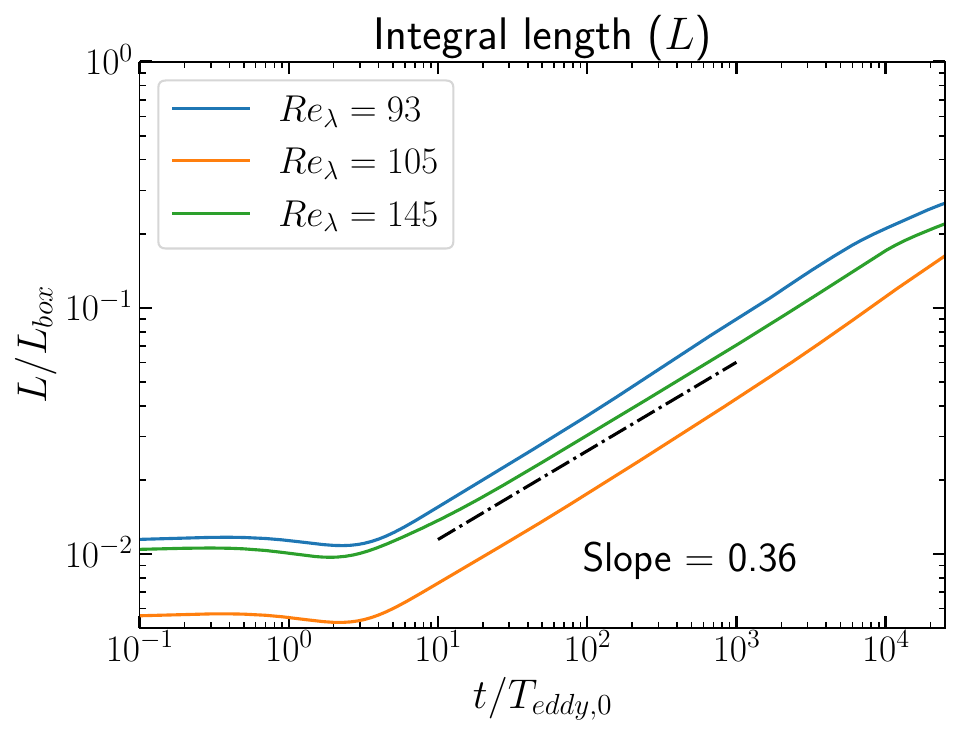}
    \caption{Time evolution of the integral length scale $L$ for simulations with an initial LKB spectrum ($E(k) \sim k^4$ for small $k$).  Different colors correspond to simulations with varying initial $Re_\lambda$, as indicated in the legend. The integral length scale on the vertical axis is normalized {by the edge length of the computational domain ($L_{box} = 2\pi$)}, and the time on the horizontal axis is normalized by the initial large-eddy turnover time, $T_{eddy,0}$. The dot-dashed line has a slope of 0.36.}
    \label{fig:int_len_k4}
\end{figure}

To understand\enlargethispage{-12pt} this breakdown, consider the following. In each case of decay, as observed already, the Reynolds number decreases only slowly through the decay period. For example, for the highest $Re_\lambda = 145$, the decrease during the entire region of constant $n$ is by a factor of 2 or so over time scales of the order of a 1000 initial integral time scales (see \hyperref[fig:power_laws_k2]{figure} \ref{fig:power_laws_k2}\hyperref[fig:power_laws_k2]{d}). We already noted that the $Re_\lambda$ at the end of the decay period could still be regarded as `high' in the sense of being far from the final period of decay.  We thus expect that the breakdown of constant $n$ is unlikely owing to a change in physics, brought about by drastic changes in the Reynolds number. On the other hand, one can see from \hyperref[fig:power_laws_k2]{figures} \ref{fig:power_laws_k2}\hyperref[fig:power_laws_k2]{a} and \ref{fig:int_len_k4} that the integral length scale $L$ increases continually, and that, roughly speaking, the regime of constant $n$ breaks down when $L$ reaches $10\%$--$15\%$ of the box size. Because the integral scale in the LKB case grows (albeit moderately) more slowly compared to the BS case, it takes significantly longer for the turbulence to feel the domain boundary. The finite-size effects are consequently delayed, extending self-similar\enlargethispage{-12pt} decay.

As a summary of the work so far, our simulations validate power laws for the kinetic energy and integral length scales, provided that the integral scale remains small relative to the domain size. Distinct power laws are observed for the BS and LKB cases, confirming that the decay dynamics are governed by the large-scale structure of the flow. In the following section, we investigate the time evolution of the full energy spectrum to explore the persistence or otherwise of spectral self-similarity. We restrict our analysis generally to $t/T_{\mathrm{eddy},0} \lesssim 2,000$, for which the flow remains effectively unconfined.

\subsection{Energy spectrum}
\label{subsec:energy_spect}
As stated earlier, the simulations were initialized using the model spectrum (equation~\ref{eq:modal_spect}) with low-wavenumber slopes of $p_0 = 2$ (for the BS case) and $p_0 = 4$ (for the LKB case). Both initializations possess a distinct $k^{-5/3}$ for higher $k$, reflecting the existence of the inertial range.

\begin{figure}
    % \centering
    \begin{subfigure}[b]{0.5\textwidth} % Adjust width as needed
        \centering
        \includegraphics[width=\linewidth]{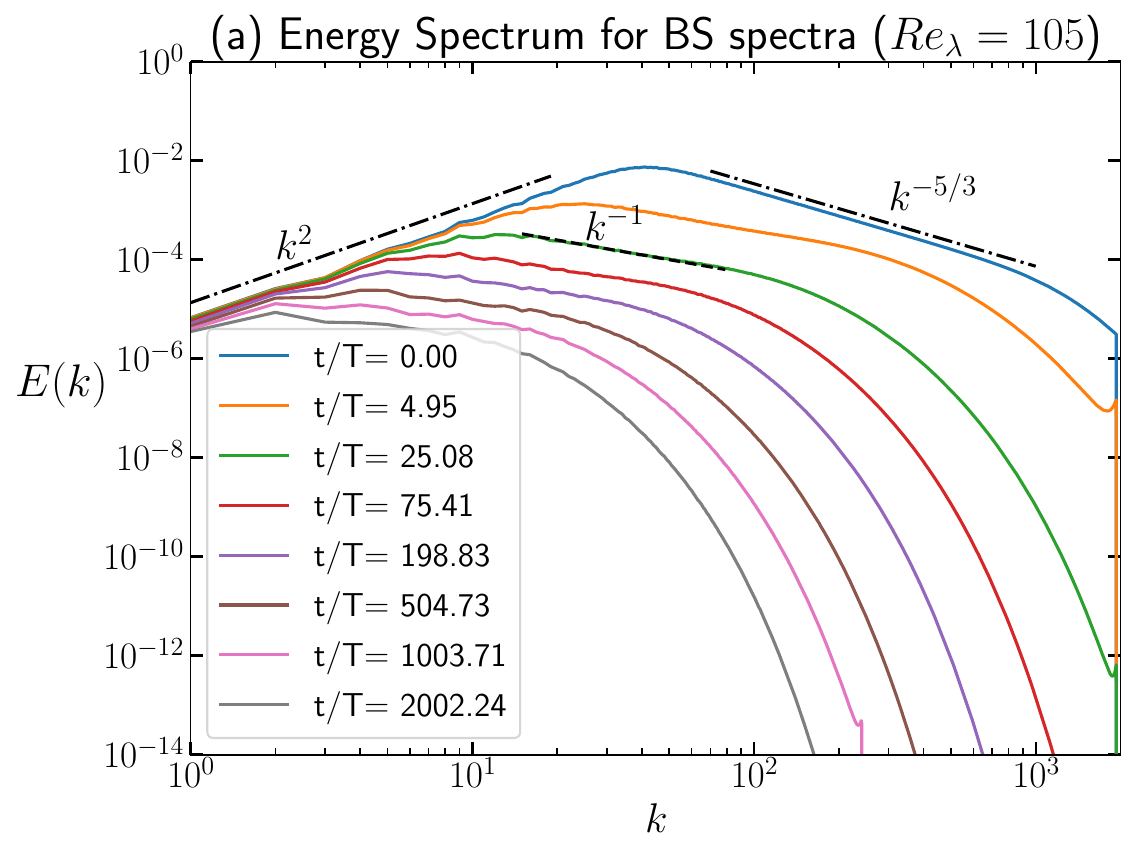}
        % \caption{}
        % \label{fig:migdal-len1}
    \end{subfigure}
    \hfill % Adds horizontal space between subfigures
    \begin{subfigure}[b]{0.5\textwidth} % Adjust width as needed
        \centering
        \includegraphics[width=\textwidth]{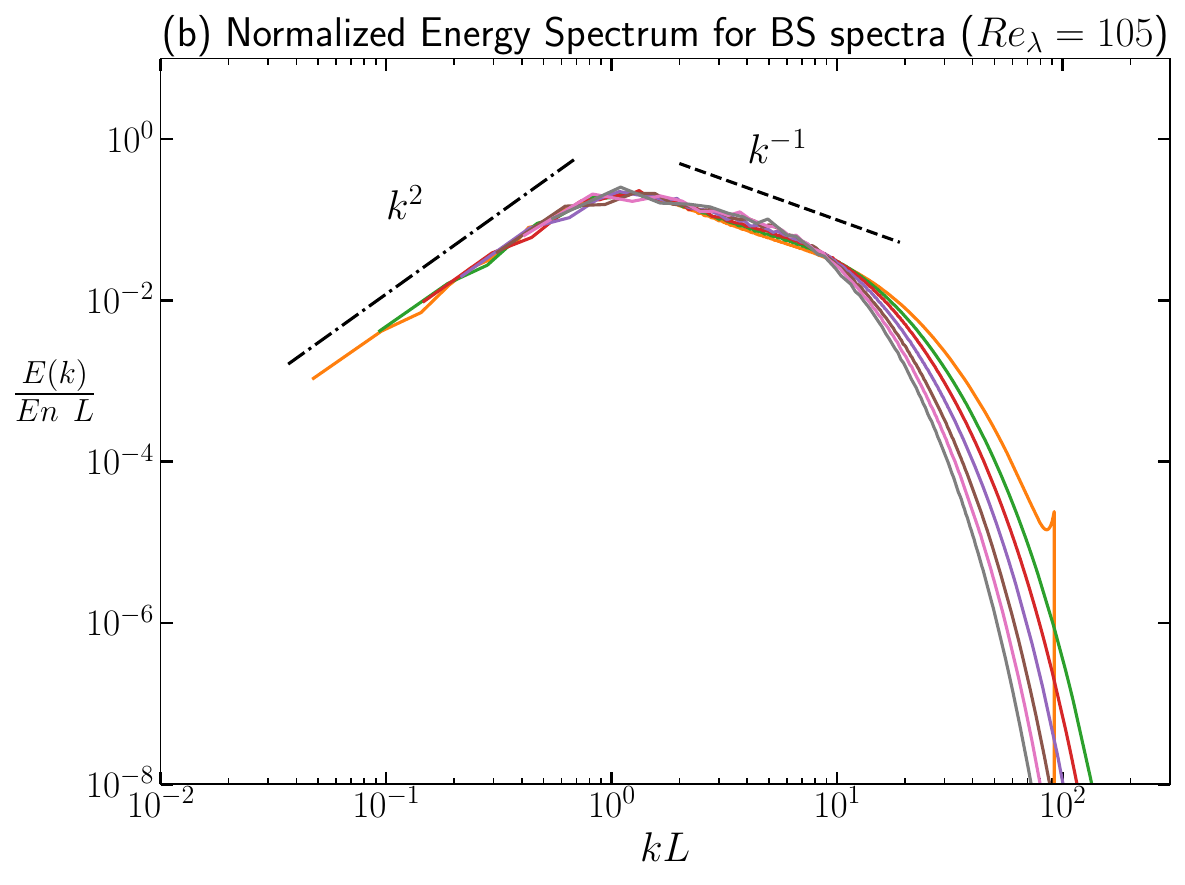} 
        % \caption{}
        % \label{fig:migdal-len1}
    \end{subfigure}
    \caption{ Time evolution of the energy spectrum $E(k)$ from a single simulation, BS case. The different colored curves represent the spectrum at various normalized times $t/T$, as indicated in the legend. (a) The uncompensated energy spectrum, $E(k)$, is plotted against the wavenumber, $k$, on log-log scales. The reference lines show that the initial $k^{-5/3}$ inertial range slope (dot-dashed) disappears quickly and yields place gradually to a $k^{-1}$ slope (dashed). (b) The energy spectrum in self-similar coordinates of $E(k)/(En~L)$ versus the normalized wavenumber $kL$  ($L$ is the time-dependent integral length scale). The data collapse from different times onto a single curve demonstrates large-scale self-similarity. The reference lines in this panel show that the initial slope of 2 (dot-dashed) for low wave numbers persists, even as the $-1$ slope  (dashed) emerges.}
    \label{fig:ene_spectra_k2}
\end{figure}

\hyperref[fig:ene_spectra_k2]{Figure}~\ref{fig:ene_spectra_k2} shows the evolution of the energy spectrum for $Re_{\lambda} = 105$ (case 5 in \hyperref[tab:sims_k2]{table}~\ref{tab:sims_k2}) at various normalized times. We selected this specific case for detailed spectral analysis because it was initialized with a particularly small value of the integral length scale ($L\approx 0.58\% \text{ of } L_{\mathrm{box}}$), resulting in an extended $k^2$ scaling range at low wavenumbers. This allows for a longer observation of the large-scale dynamics without the influence from finite-domain effects.

\begin{figure}
    % \centering
    \begin{subfigure}[b]{0.5\textwidth} % Adjust width as needed
        \centering
        \includegraphics[width=\linewidth]{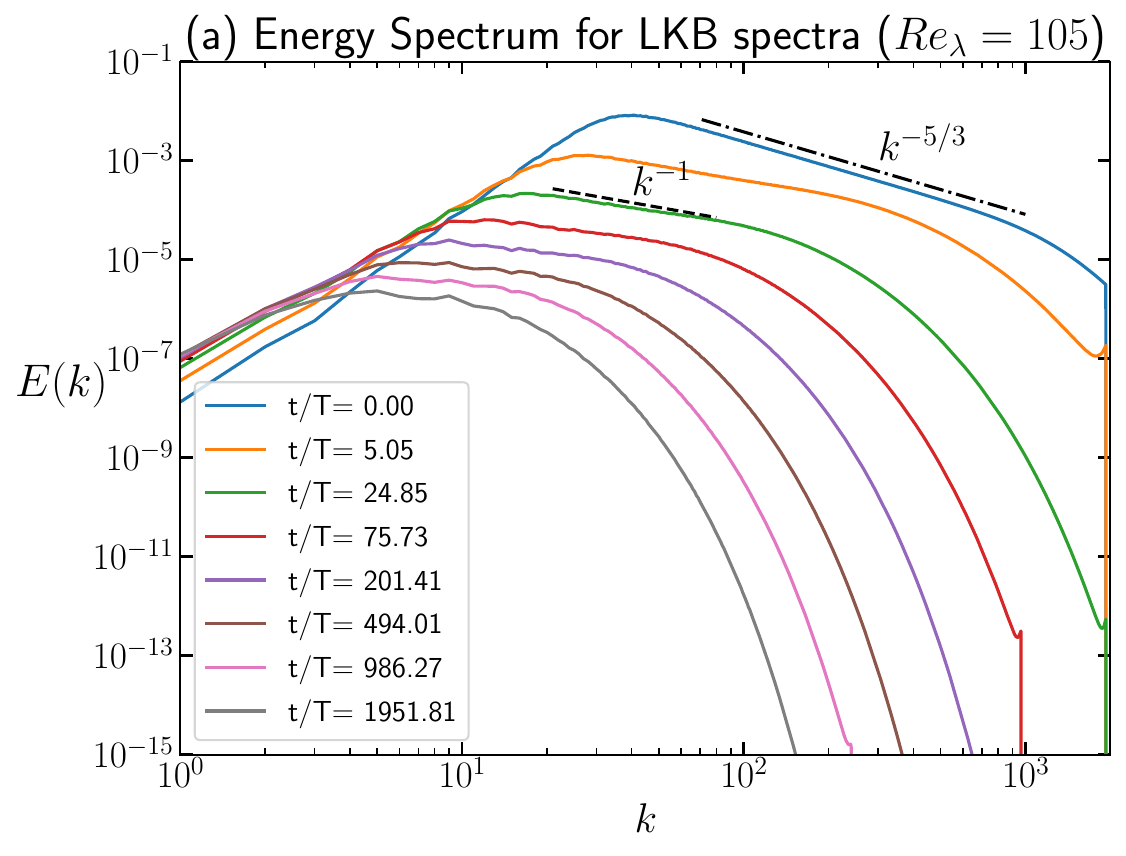}
        % \caption{}
        % \label{fig:migdal-len1}
    \end{subfigure}
    \hfill % Adds horizontal space between subfigures
    \begin{subfigure}[b]{0.5\textwidth} % Adjust width as needed
        \centering
        \includegraphics[width=\textwidth]{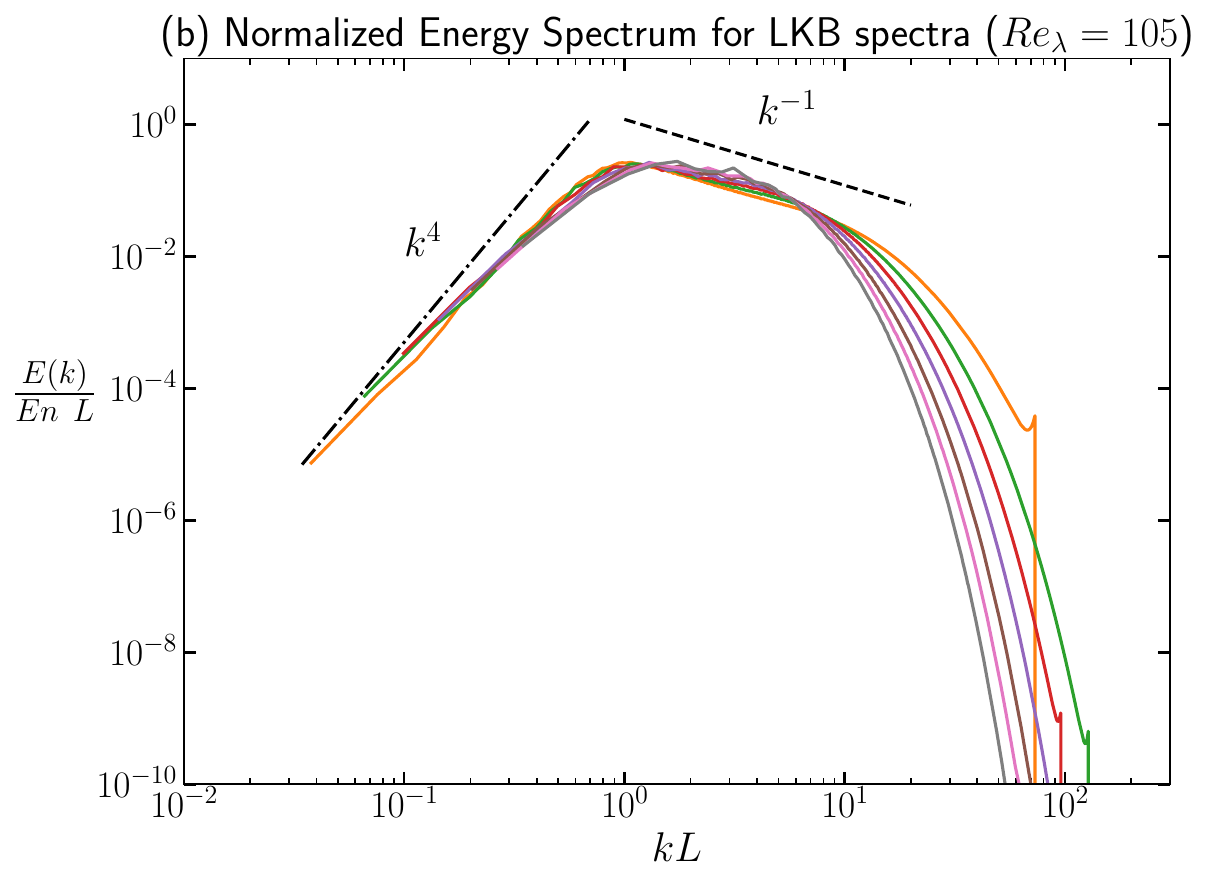} 
        % \caption{}
        % \label{fig:migdal-len1}
    \end{subfigure}
    \caption{ Time evolution of the energy spectrum $E(k)$ from a single simulation. The different colored curves in both panels represent the spectrum at various normalized times $t/T$, as indicated in the legend. (a) The uncompensated energy spectrum $E(k)$ plotted against the wavenumber $k$ on log-log coordinates. The reference lines show a $k^{-1}$ slope (dashed) and the $k^{-5/3}$ inertial range slope (dot-dashed) for comparison. (b) The energy spectrum in self-similar coordinates, plotting $E(k)/En~L$ versus the normalized wavenumber $kL$, where $L$ is the time-dependent integral length scale. The reference lines in this panel indicate a slope of -1 (dashed) and 4 (dot-dashed).} 
    \label{fig:ene_spectra_k4}
\end{figure}

\hyperref[fig:ene_spectra_k2]{Figure}~\ref{fig:ene_spectra_k2}\hyperref[fig:ene_spectra_k2]{a} displays the energy spectra. While the energy at high wavenumbers decays rapidly, the scaling at low wavenumbers ($E(k) \sim k^2$) remains valid for low wave numbers, even though it shrinks in extent as the decay proceeds: the spectral peak (initially at $k \approx 40$) shifts to lower wavenumbers. This is consistent with the principle of the permanence of large eddies; roughly speaking, it may also be interpreted to mean that the equipartition of energy holds.
A second point is the evolution of the intermediate scaling range. As shown in panel (a), the initial $k^{-5/3}$ slope is {surprisingly short-lived at all Reynolds numbers considered. (This makes us wonder if a sustained $-5/3$ spectral slope can survive in the absence of forcing.)} This scaling regime gradually gives way to an intermediate state characterized by a $k^{-1}$ slope (e.g. $t/T \ge 25$), as shown by the dashed line.
This feature is further confirmed in panel (b), where the spectra are plotted in normalized coordinates (the spectral density is normalized by the total energy $En$ and the integral length $L$, and the wavenumber by $L$): the data collapse onto a single, universal curve for $kL < 10$. The nature of self-similarity for larger wavenumbers will be discussed towards the end of this section.
For the LKB spectrum with initial $Re_\lambda = 105$ (case 2 in \hyperref[tab:sims_k4]{table}~\ref{tab:sims_k4}), selected here for its smaller initial integral length ($L\approx 0.54\% \text{ of } L_{\mathrm{box}}$), \hyperref[fig:ene_spectra_k4]{figures}~\ref{fig:ene_spectra_k4}\hyperref[fig:ene_spectra_k4]{a,b} show the unnormalized and normalized spectra, respectively. By contrast to the BS case, the $E(k) \sim k^4$ scaling does not persist: the slopes at the lowest wavenumbers decrease from the initial 4. Further, in the intermediate range, whereas the BS case transitioned to a clear $k^{-1}$ slope, the LKB case evolves to a slightly more negative slope (panel b).

\begin{figure}
    % \centering
    \begin{subfigure}[b]{0.5\textwidth} % Adjust width as needed
        \centering
        \includegraphics[width=\linewidth]{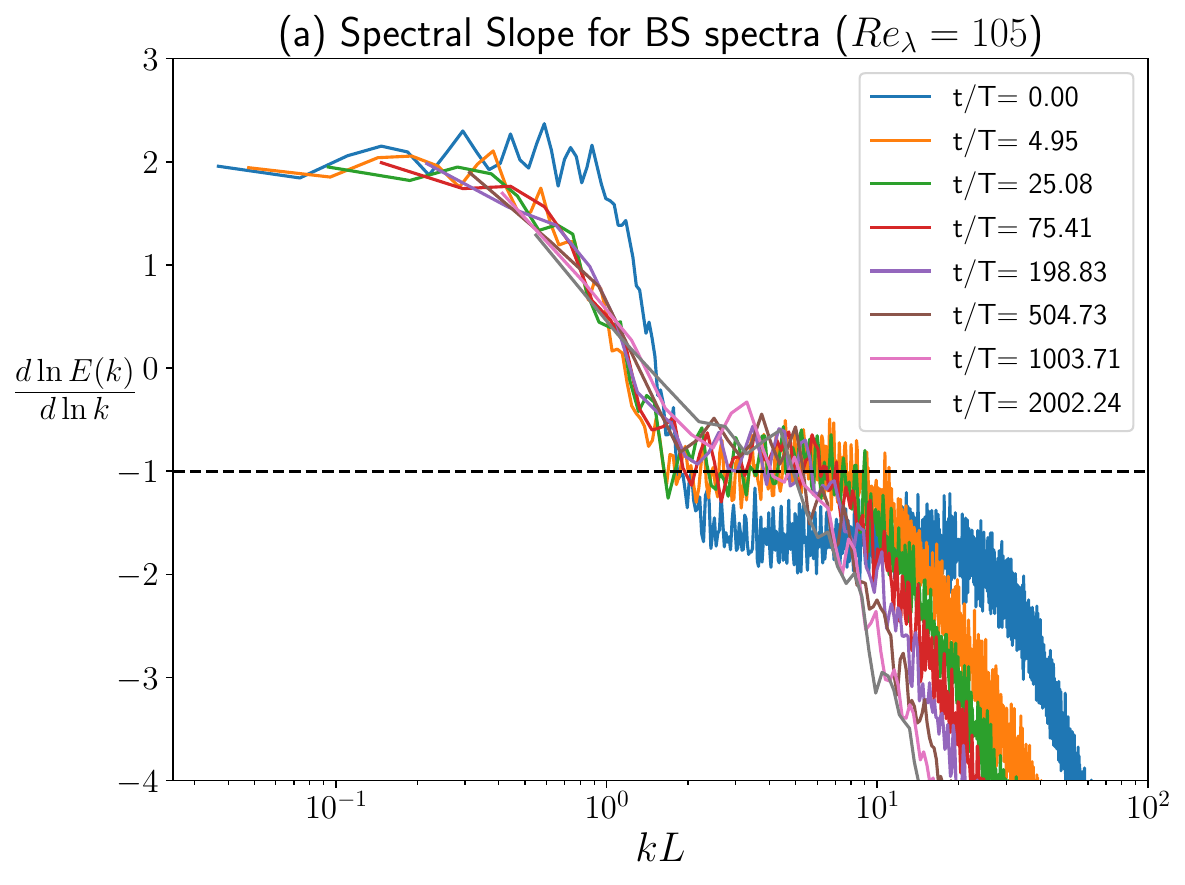}
        % \caption{Spectral slope for BS spectra ($Re_{\lambda} = 105$)}
        % \label{fig:migdal-len1}
    \end{subfigure}
    \hfill % Adds horizontal space between subfigures
    \begin{subfigure}[b]{0.5\textwidth} % Adjust width as needed
        \centering
        \includegraphics[width=\textwidth]{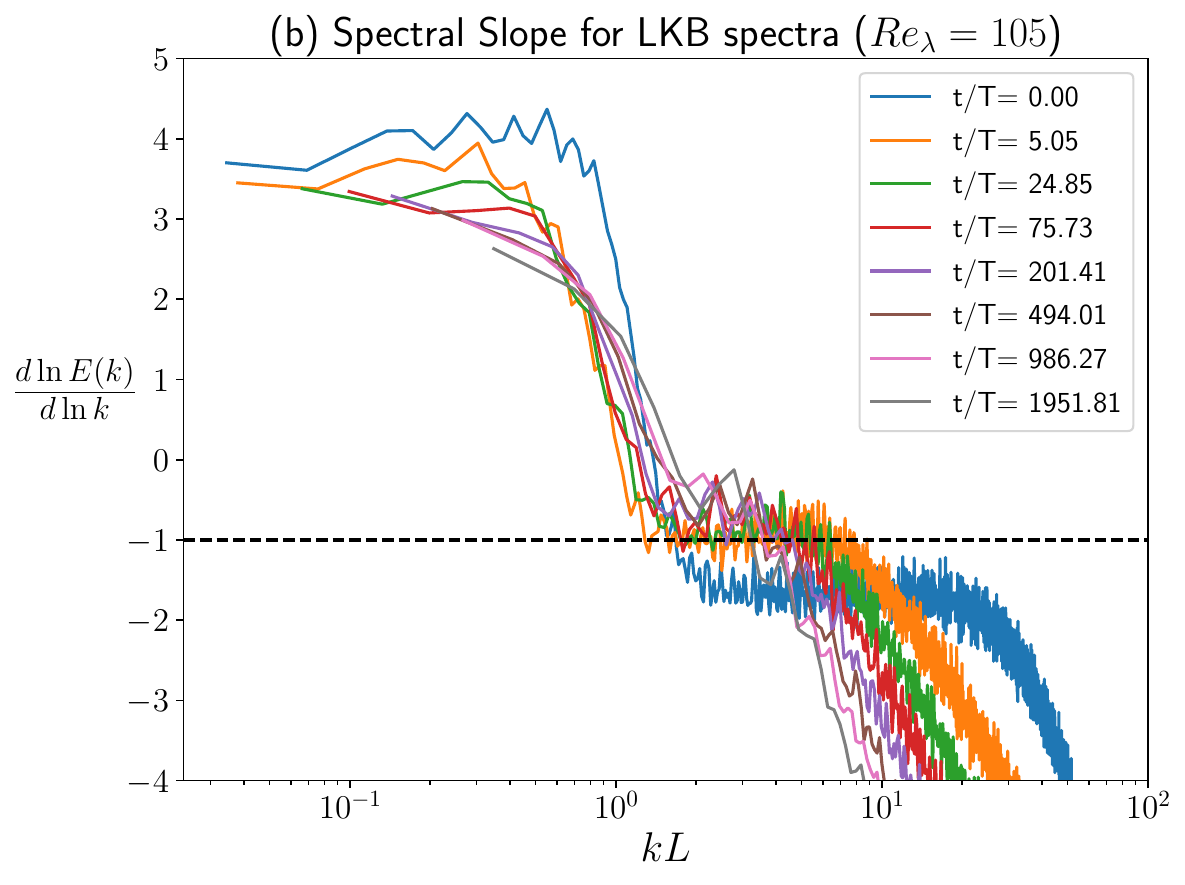} 
        % \caption{Spectral slope for LKB spectra ($Re_{\lambda} = 105$)}
        % \label{fig:migdal-len1}
    \end{subfigure}
    \caption{Time evolution of the local logarithmic slope of the energy spectrum, $d(\ln E(k)) / d(\ln k)$. The local slope is plotted against the normalized wavenumber, $kL$. The different colored curves show slopes at various normalized times $t/T$. (a) Results for the simulation with the BS spectrum. (b) Results for the simulation with the LKB spectrum. In (a), the local slope evolves to $n = -1$; this slope in (b) appears numerically to be slightly smaller.}
    \label{fig:ene_spectra_slope}
\end{figure}

\hyperref[fig:ene_spectra_slope]{Figures}~\ref{fig:ene_spectra_slope}\hyperref[fig:ene_spectra_slope]{a,b} show the local logarithmic slope of the spectrum, $d \ln E(k)/d \ln k$, for BS and LKB respectively, computed using a second-order central difference scheme and smoothed with a\break three-point moving average.
We again see that the BS results (\hyperref[fig:ene_spectra_slope]{figure}~\ref{fig:ene_spectra_slope}\hyperref[fig:ene_spectra_slope]{a}) support the principle of permanence of large eddies. In the intermediate range ($kL > 1$), the slope transitions from an initial value near $-5/3$ to fluctuate around the $-1$ reference line, confirming the $E(k) \sim k^{-1}$ scaling. On the other hand, the LKB case (\hyperref[fig:ene_spectra_slope]{figure}~\ref{fig:ene_spectra_slope}\hyperref[fig:ene_spectra_slope]{b}) shows that the low wavenumber slope departs from its initial value of 4, and assumes a slightly more negative power in the intermediate region. {A simple argument for a $-1$ power for $E(k)$ is to estimate the energy flux via the local energy spectral density and the wave number, and equate it to local dissipation $\sim k^2E(k)$. One can support this argument by numerically evaluating the energy transfer across wavenumbers, but we omit these details because they do not necessarily bring greater clarity to the argument}.

One further remark may be appropriate. It is somewhat difficult to understand why the 10/7 slope appears so robust for LKB in spite of the fact that the theoretically conditioned $k^4$ behaviour for small $k$ disappears relatively quickly in time. This is unlike the BS case for which the initial $k^2$ part remains intact (though to diminishing extents) over the duration of energy decay. In retrospect, it appears strange that the energy decay should depend so heavily on the very low number regions of the initial energy spectrum (duly noting that the initially imposed power law is turncated by the finite box size). If this effect is real (as seems to be the case), it is clear that the decay cannot be universal. If the energy decay is not universal---which may mean that, as a community, we have been chasing an ill-posed problem for very long---are there other aspects of decay that are universal? For instance, is it possible to seek universality in terms of properties that do not depend on large scales, for example, enstrophy? Can we somehow isolate the role of the `boundary effects' coming from the lowest wavenumbers? We will consider these questions in \hyperref[subsec:enstrophy]{\S\S4d} and \hyperref[sec:5]{5}, respectively, but consider comparisons with Migdal's theory first.

\newpage
\section{Comparison with Migdal's theory}
\label{sec:migdal_comp}

\subsection{The Theory}
\label{subsec:Midal_theory}
We have stated that the two classical theories designated as BS and LKB are available for comparison with the data. A preliminary comparison between them and simulations has been made in \cite{john-P-john}, and we too have already made specific comparisons at various places in this paper (with more to come). In this section, however, we highlight Migdal's theory partly because of its novelty and partly because its spirit and pedigree are different. This work is of recent vintage (2023--2025), builds on quantum field theory, and (modulo an important step) is free from semi-empirical\break closure assumptions that are common in the turbulence literature. The work combines results from quantum field theory, classical hydrodynamics, number theory etc., woven together in intriguing ways. We provide a brief description of that theory here.

The early framework for the theory is based on \cite{migdalAreaLoop95,migdalQuantLoop1995,insta_intermitent_1996,MigdalQuantumGravity_1992}, where Migdal recasts the Navier--Stokes (NS) equations into an analytically solvable form in loop space (inspired by loop quantum gravity work). Within this formalism, turbulence statistics are described not by the velocity field \textbf{v} itself, but by the circulation of velocity around a closed contour $C$,
\begin{equation}
    \Gamma_C = \oint_C \textbf{v}\cdot d\textbf{r}.
\end{equation}
The expectation value of exp~$i\Gamma_c$ defines a loop functional
\begin{equation}
    \psi[C] = \langle\exp(\mathrm{i}\Gamma_C/\nu)\rangle_t,
\end{equation}
where $\nu$ is the kinematic viscosity, {and the suffix `t' simply denotes a fixed time.} The key insight is that $\psi[C]$ satisfies a linear diffusion equation in the infinite-dimensional space of loops. This loop-space calculus transforms the nonlinear NS dynamics into a `universal' linear operator acting on the functional of loops.
This enabled Migdal to reduce the three-dimensional turbulence problem to a one-dimensional system---specifically, a quantum theory of fermions on a ring (the `Euler ensemble'---see below). This step allows turbulence statistics to be treated analytically in terms of geometrical objects such as loop derivatives and minimal surfaces, rendering the problem solvable without \textit{ad hoc} parameters.

This approach is not without problems. It is based on an ergodic hypothesis that stipulates an exact equivalence between the loop functional (the Fourier transform of the probability distribution) and the quantum trace of an evolution operator for the one-dimensional ring of Fermi particles. Bru\'e \& de Lellis \cite{BCDL} show that suitable weak solutions of NS are regular enough to make sense of the circulation over loops, allowing one to give a weak sense to the loop equation.

\subsubsection{Euler ensemble as an asymptotic solution}
The solution of Migdal's loop diffusion equation is the Euler ensemble \cite{Migdal_theory}.  {The Euler ensemble is not a stationary solution but a `universal' fixed trajectory whose time dependence leads to decaying solutions.} It represents the asymptotic state of decaying turbulence at infinite Reynolds number when the system reaches a self-similar attractor. {Despite the appearance of the word\break `Euler', it is not the solution of the Euler equations (as the appearance of $\nu$ in (4.2) clearly shows), but a WKB solution of the NS equations.} Migdal argued that the Euler ensemble is mathematically equivalent to a quantum statistical system of $N$ interacting fermions on a ring. In this formulation, each element of the turbulent flow corresponds to a discrete momentum loop $P(\theta)$, and the loop functional becomes
\begin{equation}
    \psi[C] = \left\langle \exp\left(\frac{\mathrm{i}}{\nu} \oint d\theta \dot{C} P_{\alpha}(\theta) \right)\right\rangle_P.
\end{equation}
In this one-dimensional formulation, the complex coordinate of the loop is mapped to the trajectory of fermions, providing a calculable framework for the statistical moments of the flow. A pivotal outcome of this work is the derivation of a {universal scaling function} $H(\mathcal{K})$ (often denoted as a function of a dimensionless\enlargethispage{-12pt} variable $\mathcal{K} \propto k\sqrt{\nu t}$). This function acts as the fundamental generator for the flow statistics, enabling the explicit computation of the energy decay rate, the characteristic length, and the anomalous slope of the second-order structure function.

\subsection{Key predictions for decaying turbulence}
\label{subsec:migdal_prediction}
The Euler ensemble yields explicit predictions for observable quantities in decaying turbulence. Some of them are:

\begin{itemize}
    \item[--] \textbf{Energy decay law}\\
    The turbulent kinetic energy, $En(t) = \frac{1}{2}\langle v_i v_i\rangle$, is predicted to decay asymptotically as\break $En(t)\sim t^{-5/4}$.  This result is significant as it does not assume the conservation of integral invariants, as in the BS case ($n=1.2$) or for LKB ($n = 10/7$). The Euler ensemble thus posits a universal asymptotic regime that is independent of initial conditions---a point on which we shall make explicit comments later.
    \vspace{0.2cm}
    \item[--] \textbf{The length $L_M$}\\
    Migdal introduced a characteristic length scale defined by the moments of the energy spectrum. At time $t$, it is given by
    \begin{equation}
        L_M(t) = \frac{\int kE(k,t)dk}{\int k^2E(k,t)dk},
        \label{eq:migdal_len}
    \end{equation}
    where $E(k,t)$ is the energy spectrum. Based on the asymptotic spectral shape, the theory predicts an evolution of $L_M \sim t^{1/2}$. Similar to the decay exponent, this scaling arises from the lowest mode of the Euler ensemble. This length can be computed with better control than the integral scale $L$, which depends on the large-scale behaviour of the two-point correlation length---always a source of uncertainty {in simulations as well as decaying grid turbulence}.
    \vspace{0.2cm}
    \item[--] \textbf{Slope of the second-order structure function}\\
    The theory predicts the scaling exponent for the second-order velocity structure function, defined locally as $\zeta_2(r, t) = r \partial_r \log \langle (\Delta v)^2 \rangle$. In the Euler ensemble, this depends on the normalized separation $x = r/L_M$. The theoretical curve $f(x)$ is derived via the inverse Mellin transform of the energy spectrum. Specifically, Migdal derives an explicit meromorphic function $V(p)$ (equation~96 in~\cite{Migdal_theory}) whose poles determine the spectrum of scaling dimensions.
       This function $f(x)$ is the time-averaged behaviour of $\zeta_2$.
    \vspace{0.2cm}
\item[--] \textbf{Energy spectrum at high wavenumbers}\\
    The theory predicts the form of the energy spectrum via a universal scaling function $H(\kappa)$, derived from the Euler ensemble \cite{Migdal_theory}. Asymptotically, this function is predicted to follow a power law of $H(\kappa) \sim \kappa^{-7/2}$ for large wavenumbers:
    \begin{equation}
        E(k,t) \sim k^{-7/2} \quad \text{for } k \gg 1/L_M.
    \end{equation}
    We compare this prediction with the DNS data. However, unambiguous observation of this scaling is challenging in decaying flows at moderate Reynolds numbers ($Re_{\lambda, 0} = 145$ being the maximum explored here). As the Reynolds number decays ($Re_{\lambda} \sim t^{-1/8}$), the dissipative cutoff encroaches on the inertial range, and could obscure the $k^{-7/2}$ pre-dissipative tail (if one existed) by the exponential viscous rolloff.
\end{itemize}

\subsection{Specific comparisons between Migdal's theory and simulation results}
\label{subsec:migdal_comp}

\subsubsection{The $L_M$ and energy decay $En$}
\label{subsubsec:migdal_length}
\begin{figure}
    \centering
    % BS k^2 plots
    \begin{subfigure}{0.49\textwidth}
        \includegraphics[width=\linewidth]{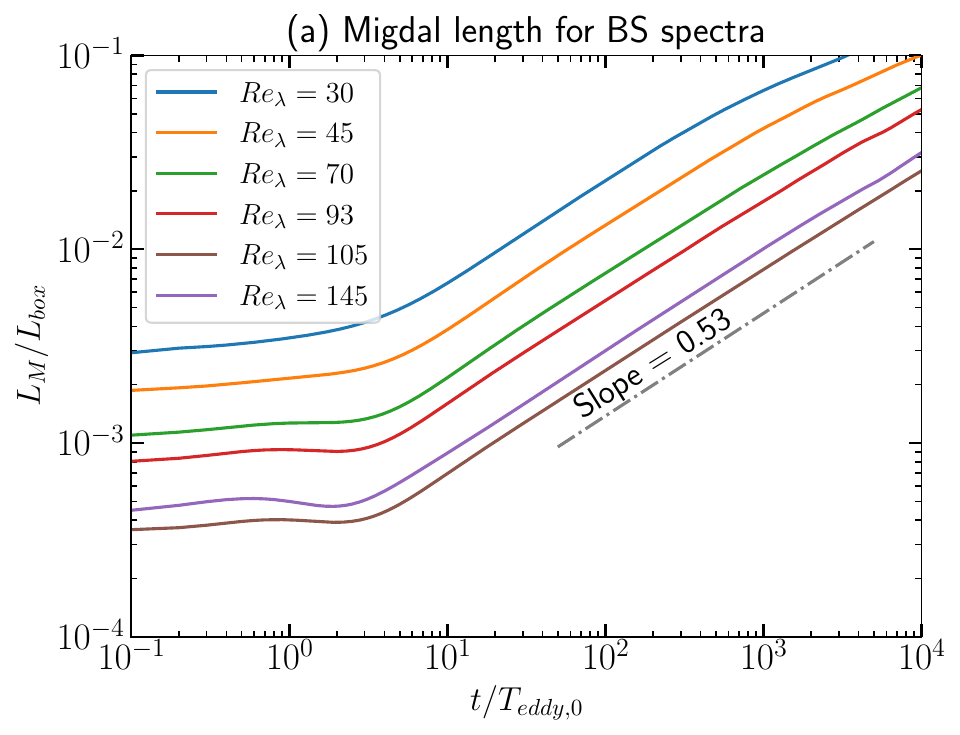} 
        % \caption{BS ($E(k) \sim k^2$) spectrum}
        % \label{fig:migdal_k2_time}
    \end{subfigure}
    \hfill
    \begin{subfigure}{0.49\textwidth}
        \includegraphics[width=\linewidth]{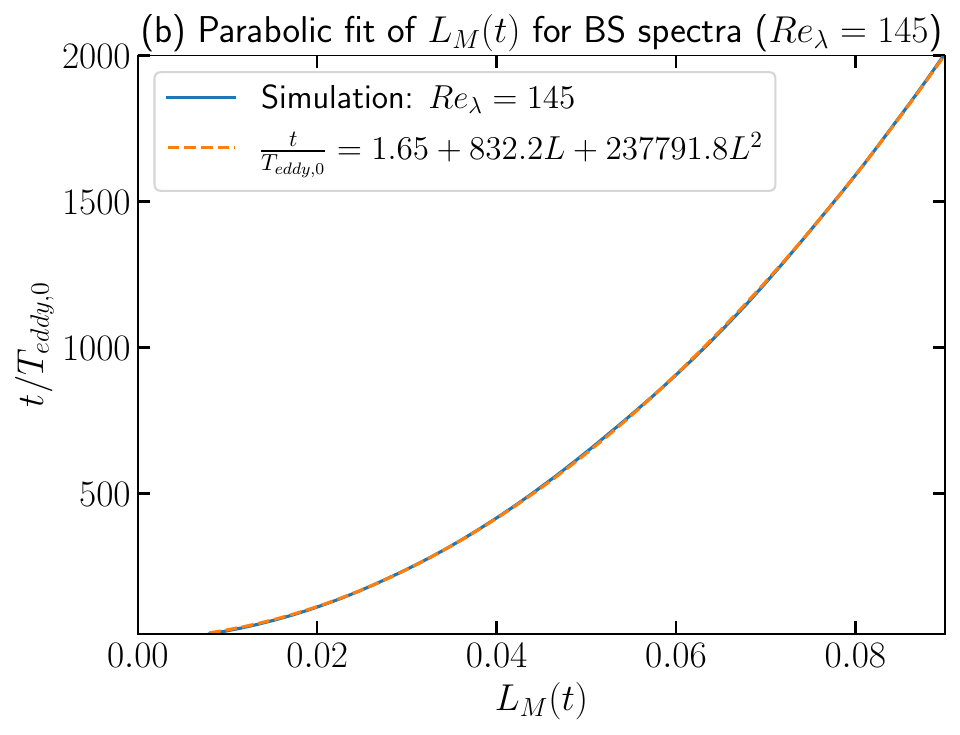}
        % \caption{BS ($E(k) \sim k^2$) spectrum, $Re_\lambda=145$}
        % \label{fig:migdal_k2_fit}
    \end{subfigure}
    
    % \vspace{0.2cm} % Optional vertical spacing between rows
    
    % LKB k^4 plots
    \begin{subfigure}{0.49\textwidth}
        \includegraphics[width=\linewidth]{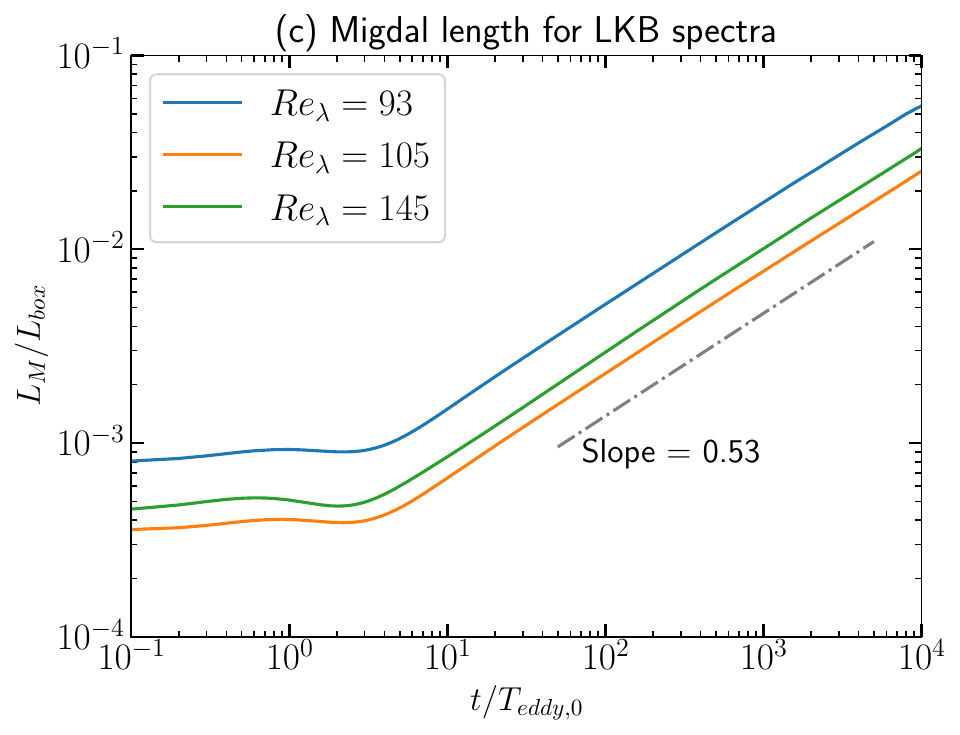} % <--- Put your LKB L_M vs t plot file here
        % \caption{LKB ($E(k) \sim k^4$) spectrum}
        % \label{fig:migdal_k4_time}
    \end{subfigure}
    \hfill
    \begin{subfigure}{0.49\textwidth}
        \includegraphics[width=\linewidth]{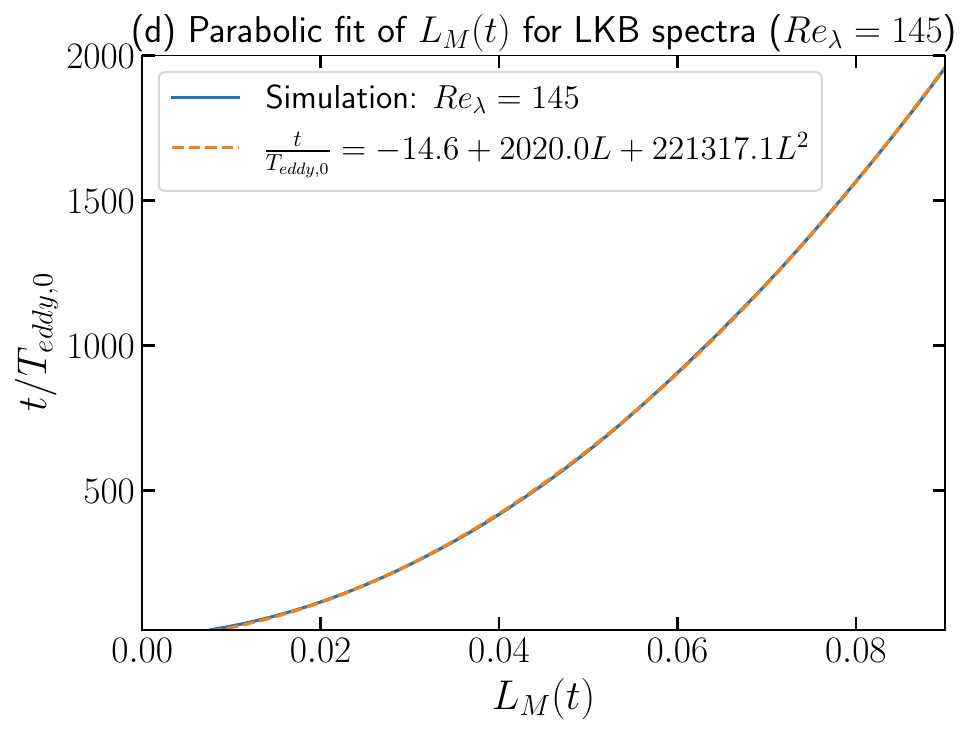} % <--- Put your LKB parabola fit plot file here
        % \caption{LKB ($E(k) \sim k^4$) spectrum, $Re_\lambda=145$}
        % \label{fig:migdal_k4_fit}
    \end{subfigure}

    \caption{The Migdal length $L_M$, defined in Eq.~(\ref{eq:migdal_len}), for simulations with BS ($E(k)\sim k^2$ for $k \to 0$) and LKB ($E(k)\sim k^4$ for $k \to 0$) initial spectra.
    \textbf{(a, c):} Time evolution of $L_M$ normalized by the box length, $L_{box}$, for (a) the BS and (c) the LKB cases. The best fit over the region of constant $n$ yields $L_M \sim t^{0.53}$, slightly higher than the theoretical $t^{0.5}$.
    \textbf{(b, d):} Time $t$ plotted as a function of $L_M(t)$ for the $Re_\lambda=145$ simulations, confirming the parabolic relationship $t \sim L_M^2$. The dashed lines show the quadratic fits:
    (b) $t=1.65 + 832.2L_M + 237791.8L_M^2$ for the BS case.
    (d) $t=-14.6 + 2020.0L_M + 221317.1L_M^2$ for the LKB case.}
    \label{fig:migdal_length_combined}
\end{figure}

The length scale $L_M$ is predicted to vary asymptotically as $L_M \sim \sqrt{t}$, which we assess in \hyperref[fig:migdal_length_combined]{figure}~\ref{fig:migdal_length_combined}. Panel (a) shows the time evolution of $L_M$ for all BS cases (see \hyperref[tab:sims_k2]{table}~\ref{tab:sims_k2}), normalized by the box length ($L_{\mathrm{box}} = 2\pi$) and the initial eddy turnover time $T_{\mathrm{eddy},0}$.
$L_M$ grows with a power-law exponent of approximately $0.53$, which is close to the theoretical value of $0.5$.
If we fit time as a parabolic function of $L_M$, allowing for a virtual origin, we obtain excellent fits. Panel (b) plots $t$ against $L_M$ for Case 6 ($Re_{\lambda}=145$), along with the fitted curve $t=1.65 + 832.2L_M + 237791.8L_M^2$. This confirms that the growth is predominantly quadratic ($t \sim L_M^2$), with the linear term accounting for the virtual origin of the length scale as a subleading correction.
We performed the same analysis for the LKB simulations. Panel (c) shows a comparable power law $L_M \sim t^{0.53}$ after an initial transient of $\sim 10 T_{\mathrm{eddy},0}$. Panel (d) compares the case 3 simulation ($Re_{\lambda}=145$) with its fitted parabola, $t=-14.6 + 2020.0L_M + 221317.1L_M^2$. For both cases, the parabolic relationship confirms that $L_M$ is a more robust length scale capturing self-similar decay than the standard integral scale (see \hyperref[{fig:power_laws_k2}]{figure}~\ref{fig:power_laws_k2}).

\begin{figure}
    % \centering
    \begin{subfigure}[b]{0.5\textwidth} % Adjust width as needed
        \centering
        \includegraphics[width=\linewidth]{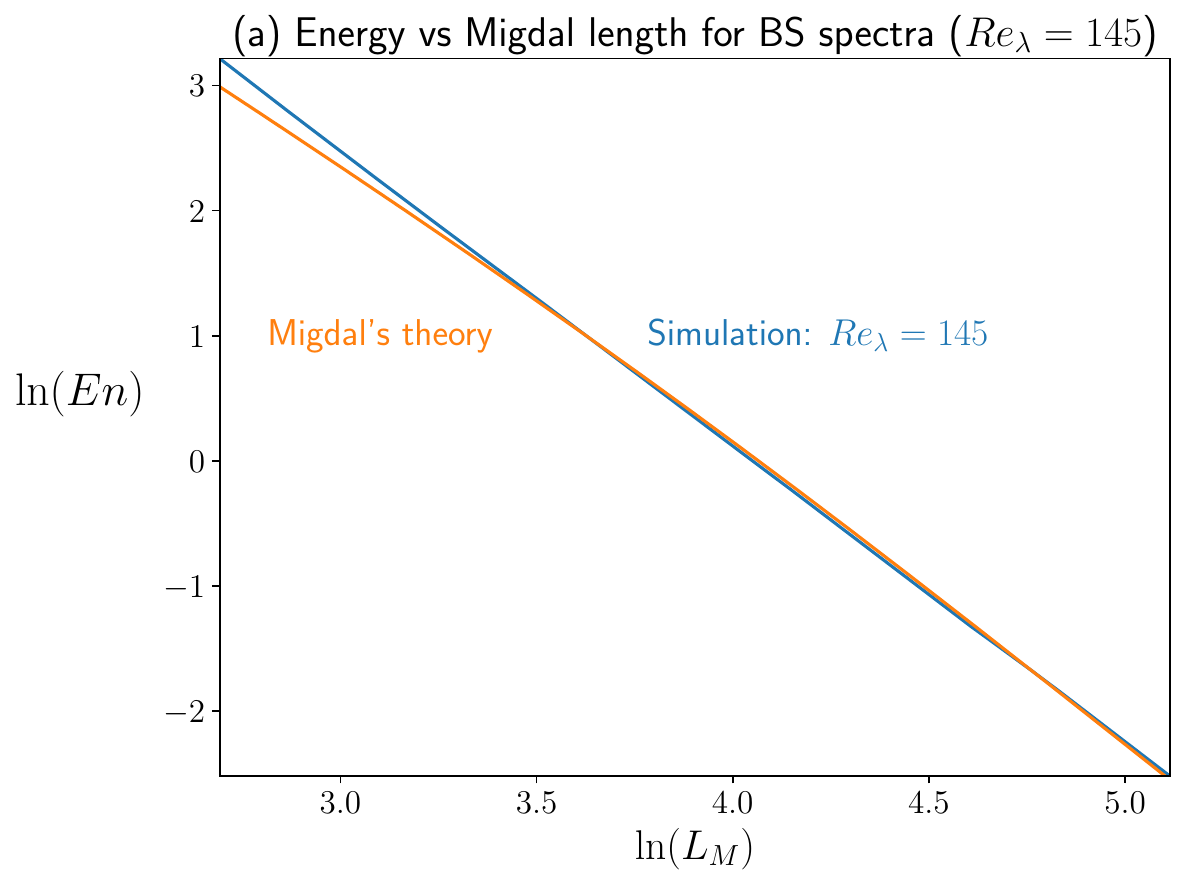}
        % \caption{}
        % \label{fig:Migdal_len_ene_k2}
    \end{subfigure}
    \hfill % Adds horizontal space between subfigures
    \begin{subfigure}[b]{0.5\textwidth} % Adjust width as needed
        \centering
        \includegraphics[width=\linewidth]{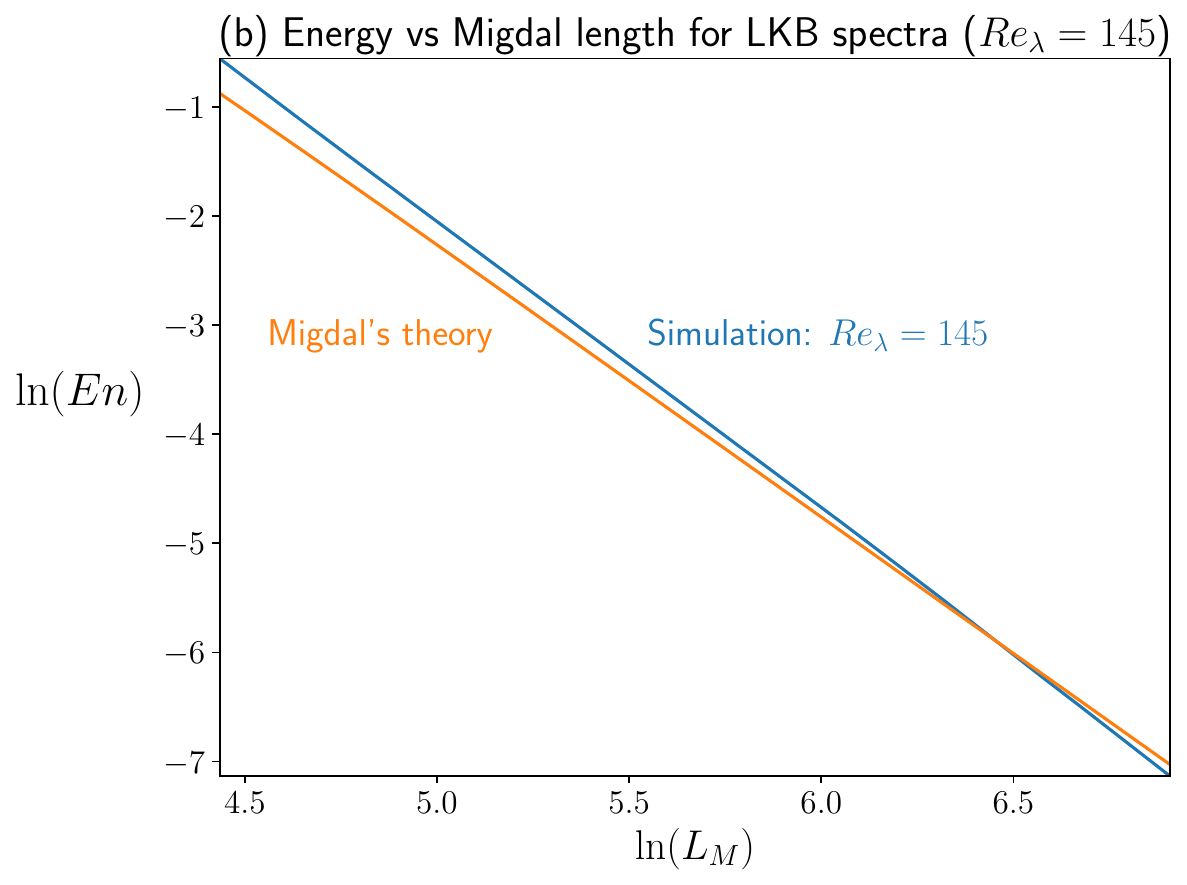}
        % \caption{}
        % \label{fig:Migdal_len_ene_k4}
    \end{subfigure}
    \caption{The relationship between total kinetic energy $En$ and the length $L_M$. The plots show the natural logarithm of energy, $\ln En$, as a function of $\ln L_M$. (a) Results for the BS ($E(k) \sim k^2$) spectrum simulation at $Re_\lambda = 145$. (b) Results for the LKB ($E(k) \sim k^4$) spectrum simulation at $Re_\lambda = 145$. In both panels, the simulation data (solid blue line) is compared to the theoretical prediction (solid orange line), showing poorer agreement for the LKB case.}
    \label{fig:migdal_len_ene}
\end{figure}

To test the power--law relationship predicted by the theory differently, we plot the logarithm of energy against log $L_M$ in \hyperref[fig:migdal_len_ene]{figure}~\ref{fig:migdal_len_ene}. Panels (a) and (b) compare the theoretical prediction with the BS (case 6) and LKB (case 3) simulations, respectively.
Migdal's theory, based on the asymptotic scalings $En \sim t^{-5/4}$ and $L_M \sim t^{1/2}$, predicts a specific slope of $-2.5$ for the $\ln En$ versus $\ln L_M$ curve. For the BS case in panel (a), the agreement is excellent sufficiently long after the onset of decay.
By contrast, the LKB case (panel b) is a significant deviation from the theoretical prediction: one observes a steeper slope of $\approx -2.7$ in (b), measurably different from $-2.5$.
This discrepancy highlights a limitation in the universality claimed for the Euler ensemble framework (but see comments at the end of \hyperref[subsec:energy_spect]{\S3b} {on the very existence of universality of energy decay}). Although the theory purports to provide a complete solution for decaying turbulence \cite{Migdal_theory}, our results suggest that it works well for the BS case but does not apply to the LKB regime. We will return to this point momentarily.

\subsubsection{Slope of second-order structure function}
\label{subsubsec:zeta2_fx}
\begin{figure}
    % \centering
    \begin{subfigure}[b]{0.5\textwidth} % Adjust width as needed
        \centering
        \includegraphics[width=\linewidth]{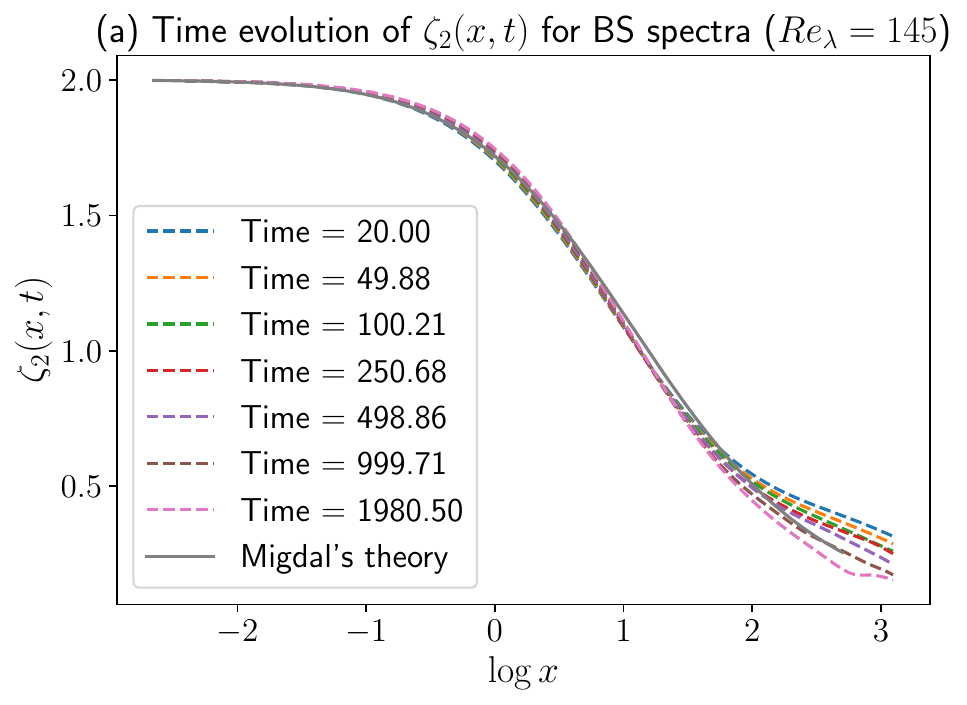}
        % \caption{}
        % \label{fig:zeta2}
    \end{subfigure}
    \hfill % Adds horizontal space between subfigures
    \begin{subfigure}[b]{0.5\textwidth} % Adjust width as needed
        \centering
        \includegraphics[width=\linewidth]{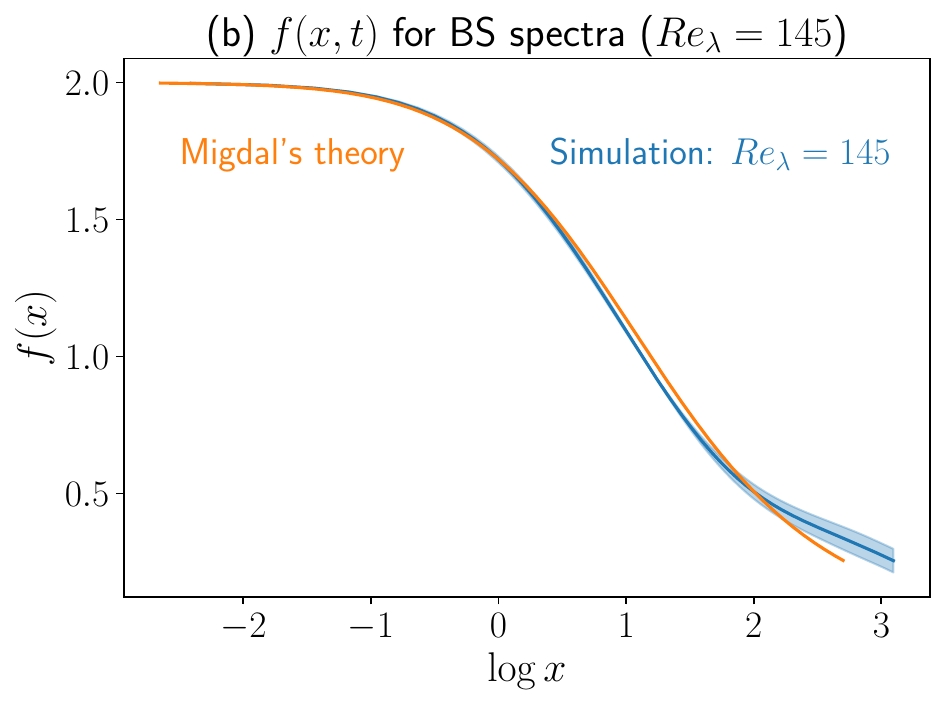}
        % \caption{f(x) for BS spectrum}
        % \label{fig:f(x)_k2}
    \end{subfigure}
    \caption{The local scaling exponent of the second-order structure function, $\zeta_2(x, t)$. This exponent is defined as $\zeta_2(x, t) = r \partial_r \log(\langle \Delta v^2 \rangle)(r)$, where $x$ is the separation distance $r$ normalized by $L_M$ (e.g., $x = r/L_M(t)$). (a) The evolution of $\zeta_2(x, t)$ at various normalized times $t/T$ (dashed lines) for the $Re_\lambda = 145$ simulation. These are compared to the solid black line representing the prediction from the theory. The collapse of the simulation data onto a single curve demonstrates self-similarity. (b) The time-averaged local exponent, $f(x) = \langle \zeta_2(x, t) \rangle_t$, for the $Re_\lambda = 145$ simulation (solid blue line), compared with Migdal's theory (solid orange line). The shaded area around the simulation curve represents the temporal standard deviation.}
    \label{fig:zeta_and_mean_k2}
\end{figure}

\begin{figure}
    % \centering
    \begin{subfigure}[b]{0.5\textwidth} % Adjust width as needed
        \centering
        \includegraphics[width=\linewidth]{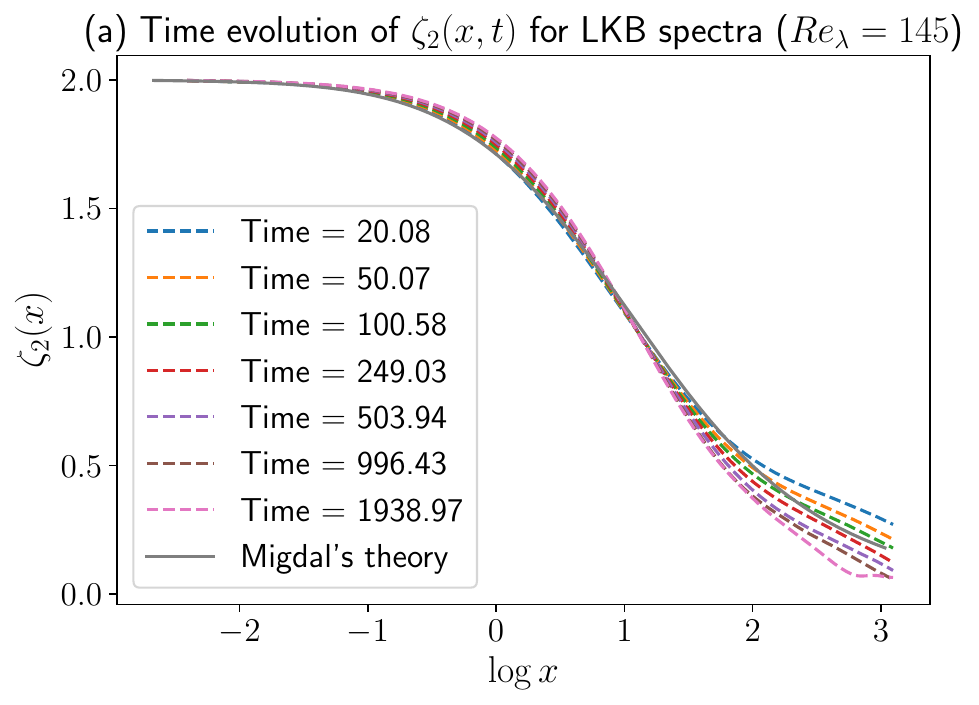}
        % \caption{}
        % \label{fig:zeta2_k4}
    \end{subfigure}
    \hfill % Adds horizontal space between subfigures
    \begin{subfigure}[b]{0.5\textwidth} % Adjust width as needed
        \centering
        \includegraphics[width=\linewidth]{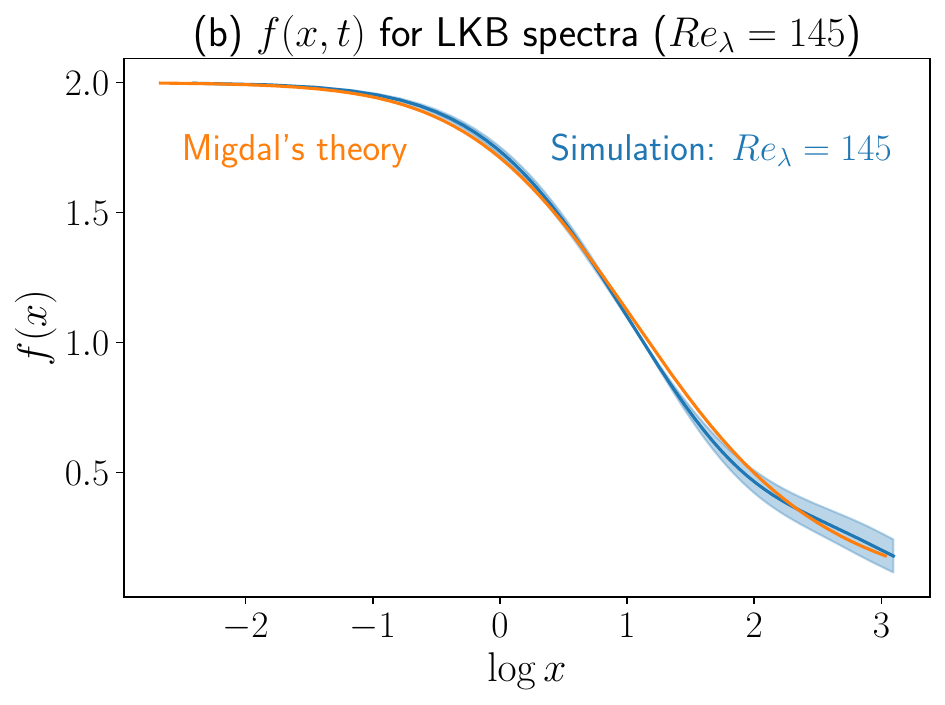}
        % \caption{}
        % \label{fig:f(x)_k4}
    \end{subfigure}
    \caption{The local scaling exponent $\zeta_2(x, t)$ for simulations with an initial LKB spectrum ($E(k) \sim k^4$). The exponent is defined as $\zeta_2(x, t) = r \partial_r \log(\langle \Delta v^2 \rangle)(r)$, plotted against the normalized separation $x$. (a) The evolution of $\zeta_2(x, t)$ at various normalized times $t/T$ (dashed lines). The collapse of the data demonstrates self-similarity. The solid black line shows the prediction from Migdal's theory for comparison. (b) The time-averaged local exponent, $f(x) = \langle \zeta_2(x, t) \rangle_t$, for the $Re_\lambda = 145$ simulation (solid blue line), compared with Migdal's theory (solid orange line). The shaded area represents the temporal standard deviation of the simulation data.}
    \label{fig:zeta_and_mean_k4}
\end{figure}

We now examine the internal structure of the flow using the local scaling exponent of the second-order structure function, $\zeta_2$. As outlined above, Migdal's theory predicts that $\zeta_2(r,t) = r \partial_r \ln \langle (\Delta v)^2 \rangle$ is a universal function $f(x)$ of the normalized separation $x = r/L_M$.
\hyperref[fig:zeta_and_mean_k2]{Figure}~\ref{fig:zeta_and_mean_k2} plots $\zeta_2$ against $\ln(x)$ for the BS case ($Re_{\lambda}=145$). Panel (a) shows the evolution at various normalized times, while panel (b) shows the time-averaged behaviour over the power-law regime ($20 \le t/T_{\mathrm{eddy},0} \le 2000$). In panel (a), the simulation curves exhibit an excellent collapse onto the theoretical prediction for $\ln(x) < 1.5$. At larger separations, the curves bend downwards, perhaps reflecting the finite size of the computational domain which naturally truncates the inertial correlations. The time-averaged curve in panel (b) confirms this agreement, with the shaded region indicating minimal temporal fluctuations around the theoretical curve $f(x)$.
A parallel analysis for the LKB case, shown\break in \hyperref[fig:zeta_and_mean_k4]{figure}~\ref{fig:zeta_and_mean_k4}, shows striking correspondence with the theory despite the departure observed in the energy decay rates.  This suggests that the self-similar spectral shape predicted by the Euler ensemble is a robust feature of decaying turbulence, persisting even when the global decay rate is different from the prediction.

\subsubsection{Energy spectra at high-wavenumber range}
\label{subsubsec:ene_spect}
\begin{figure}
    \centering
    % Left: Compensated spectrum (k^-5/3)
    \begin{subfigure}{0.49\textwidth}
        \centering
        \includegraphics[width=\linewidth]{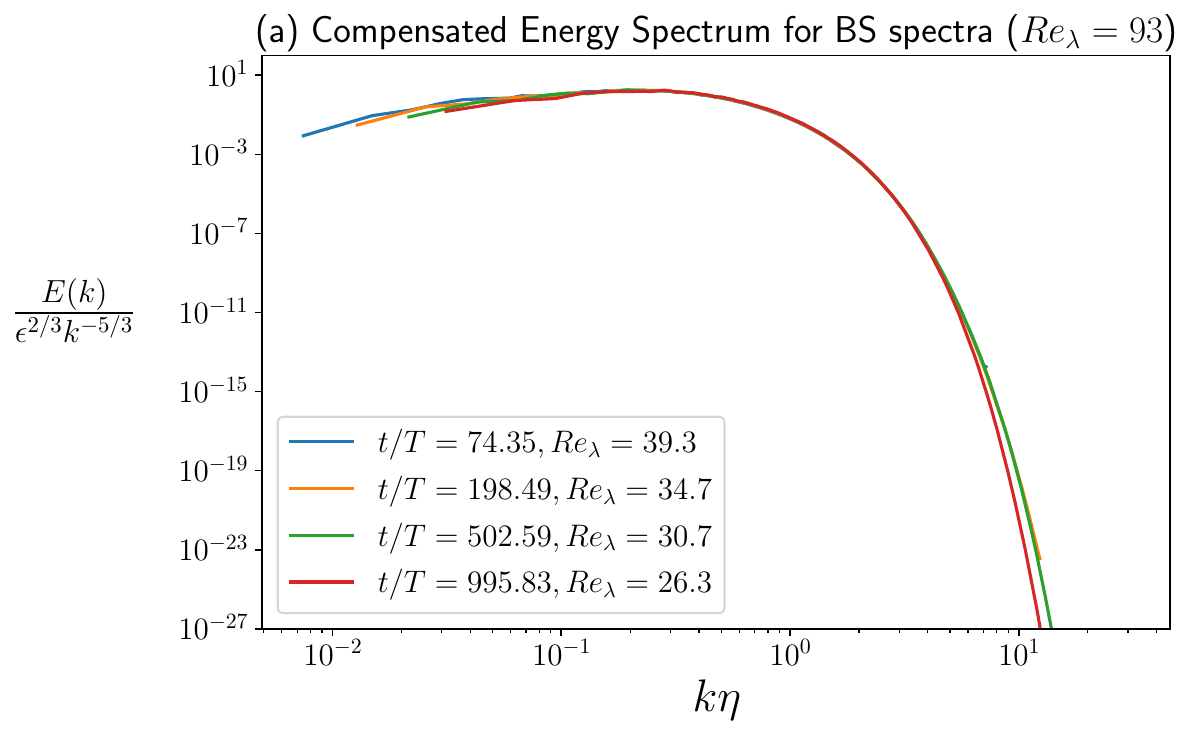} % Inertial Range Compensated Spectrum
        % \caption{Compensated Spectrum ($E(k)/k^{-5/3}$)}
        % \label{fig:diss_spec_comp}
    \end{subfigure}
    \hfill
    % Right: Local Slope
    \begin{subfigure}{0.49\textwidth}
        \centering
        \includegraphics[width=\linewidth]{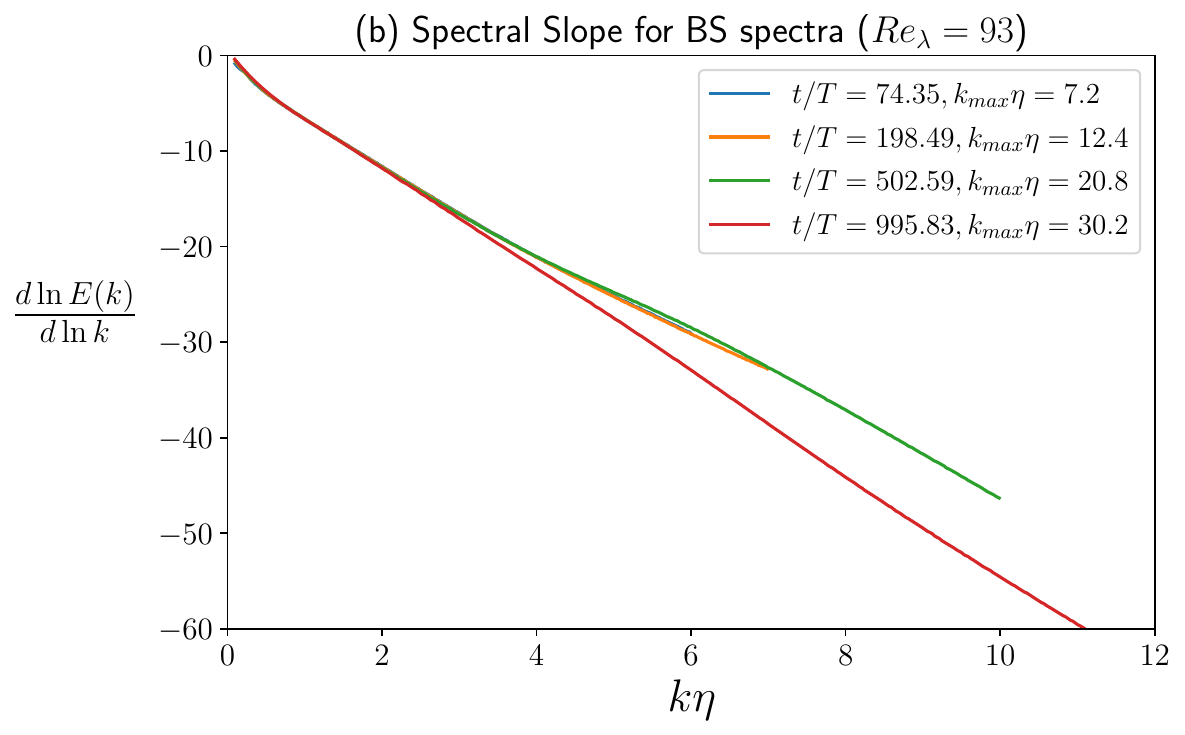} % Local Slope
        % \caption{Local Slope}
        % \label{fig:diss_spec_slope}
    \end{subfigure}
    
    \caption{Spectral statistics in the dissipation range for the $Re_{\lambda} = 93$ simulation ($2048^3$, no grid modification). (a) Energy spectra compensated by the inertial range scaling $\epsilon^{2/3}k^{-5/3}$. The lack of a horizontal plateau reflects the low Reynolds number, but the high-wavenumber collapse is robust. (b) The local logarithmic slope of the energy spectrum. The slope decreases monotonically, consistent with a stretched-exponential decay. The collapse of the curves across different times ($Re_\lambda$ decreasing from 39.3 to 26.3) demonstrates the robustness of the dissipation range scaling.}
    \label{fig:dissipation_spectra}
\end{figure}

We investigate the spectral behaviour at high wavenumbers for the $2048^3$ simulation with initial $Re_{\lambda} = 93$ (case 4, \hyperref[tab:sims_k2]{table}~\ref{tab:sims_k2}). This\enlargethispage{-12pt} run was performed without grid modification to ensure that small-scale statistics are free from minor artefacts of interpolation and truncation near the spectral cutoff. \hyperref[fig:dissipation_spectra]{Figure}~\ref{fig:dissipation_spectra}\hyperref[fig:dissipation_spectra]{a} shows the energy spectrum compensated by the inertial range scaling, $E(k)k^{5/3}\epsilon^{-2/3}$. At these moderate Reynolds numbers, a discernible $k^{-5/3}$ plateau is not observed; instead, the\break compensated spectrum exhibits a peak before dropping off into the dissipation range. The curves at high wavenumbers collapse for $k\eta < 3$ or so, for $Re_\lambda$ decreasing from 39.3 to 26.3.

\hyperref[fig:dissipation_spectra]{Figure}~\ref{fig:dissipation_spectra}\hyperref[fig:dissipation_spectra]{b} presents the local logarithmic slope of the energy spectrum, defined as
\begin{equation}
    n(k) = \frac{d \ln E(k)}{d \ln k}.
    \label{eq:local_slope}
\end{equation}
We fit the data to the functional form proposed by Buaria et al. \cite{Buaria2020}, given by
\begin{equation}
    \frac{d \ln E(k)}{d \ln k} = \alpha - \beta \gamma (k\eta)^{\gamma},
    \label{eq:stretched_exp_slope}
\end{equation}
which corresponds to a spectrum of the form $E(k) \sim k^\alpha \exp(-\beta (k\eta)^\gamma)$. The fitted parameters for different times are summarized in Table~\ref{tab:slope_params}.

Consistent with recent high-resolution studies \cite{Sualeh_2018,Buaria2020}, we do not observe pure power law. We do observe, however, a remarkable collapse of the slopes in the near-dissipation range, specifically for $0.2 < k\eta < 3$, at all times. However, in the far-dissipation range ($k\eta > 3$), the behaviour is different. For the earlier times ($t/T = 74.35$ and $198.49$), the slope decreases slowly (and non-linearly). This behaviour is captured by the fitting parameter $\gamma$ in \hyperref[tab:slope_params]{table}~\ref{tab:slope_params}, which takes values of {$8.266\times 10^{-1}$} and {$8.225\times 10^{-1}$}, respectively. A value of $\gamma < 1$ corresponds to a stretched-exponential decay, consistent with forms often cited in high-Reynolds-number turbulence. In contrast, at later times ($t/T \approx 500$ and 1000), where the Reynolds number has decreased further ($Re_{\lambda} \approx 30.7$ and $26.3$), the parameter $\gamma$ converges to approximately unity ({$\gamma = 1.009$} and {$1.019$ for the two cases).} With $\gamma \approx 1$, equation~\ref{eq:stretched_exp_slope} reduces to a linear function of wavenumber, $n(k) \approx \alpha - \beta (k\eta)$, confirming the visual observation of a constant linear rate of decrease. This distinct shift indicates that the functional form of the viscous cutoff evolves as the Reynolds number drops, transitioning towards a simpler decay scaling in the deep viscous range as the flow laminarizes.

The agreement with\enlargethispage{-12pt} this model suggests that the small-scale statistics in decaying turbulence share the same universality class as forced turbulence in the far-dissipation range, governed by the analytic properties of the NS equations, despite the differences in large-scale forcing. It should be pointed out in passing that the authors of \cite{eyink} have resurrected the view that very small scales of turbulence cannot be understood by a study of deterministic NS equations because of the considerable influence of stochastic forcing by molecular motions. We are sympathetic to this view in principle but agnostic to it in many instances encountered in practice, and present our results as we find within the present scope.

\begin{table}[h]
\centering
\begin{tabular}{|c|c|c|c|}
\hline
$t/T_{eddy,0}$ & $\alpha$  & $\beta$  & $\gamma$ \\ \hline
$7.435\times 10^{1}$ & $3.274\times 10^{-1}$ & $8.209 \times 10^{0}$ & $8.266\times 10^{-1}$ \\ \hline
$1.985\times 10^{2}$ & $3.447\times 10^{-1}$ & $8.265 \times 10^{0}$ & $8.225\times 10^{-1}$ \\ \hline
$5.026\times 10^{2}$ & $-3.408\times 10^{-1}$ & $4.143 \times 10^{0}$ & $1.009\times 10^{0}$ \\ \hline
$9.958\times 10^{2}$ & $-1.434\times 10^{-1}$ & $4.998 \times 10^{0}$ & $1.019\times 10^{0}$ \\ \hline
\end{tabular}
\caption{Fitting parameters for the local spectral slope using the model in Eq.~(\ref{eq:stretched_exp_slope}).}
\label{tab:slope_params}
\end{table}

As a summary of this part of the work, the comparison between the present well-resolved DNS data and Migdal's theory reveals a nuanced picture. For the BS regime, the agreement is\break comprehensive: the square-root growth of $L_M$, the energy decay trajectory ($\ln En$ vs. $\ln L_M$), and the internal spectral structure ($\zeta_2$) all closely follow theoretical predictions based on the $n=5/4$ law.
The\enlargethispage{-12pt} LKB regime presents a somewhat mixed message. The global energy decay ($n\approx 10/7$) explicitly departs from the theory's $n=5/4$ baseline. However, the theory remains predictive of the internal structure of turbulence. The evolution of the length scale and the universal shape of the structure function ($\zeta_2$) are captured accurately. This may suggest that while the rate of decay is sensitive to large-scale initial conditions, the distribution of energy across the inertial range relaxes to the universal attractor described by the theory.
Finally, in the high wavenumber range, we do not observe the predicted $k^{-7/2}$ spectral tail; if one exists, it could be obscured by the exponential viscous rolloff at the Reynolds numbers of this study. It is apparent that studies at higher Reynolds numbers are called for.

\subsection{Decay of enstrophy}
\label{subsec:enstrophy}
\begin{figure}
    % \centering
    \begin{subfigure}[b]{0.5\textwidth} % Adjust width as needed
        \centering
        \includegraphics[width=\linewidth]{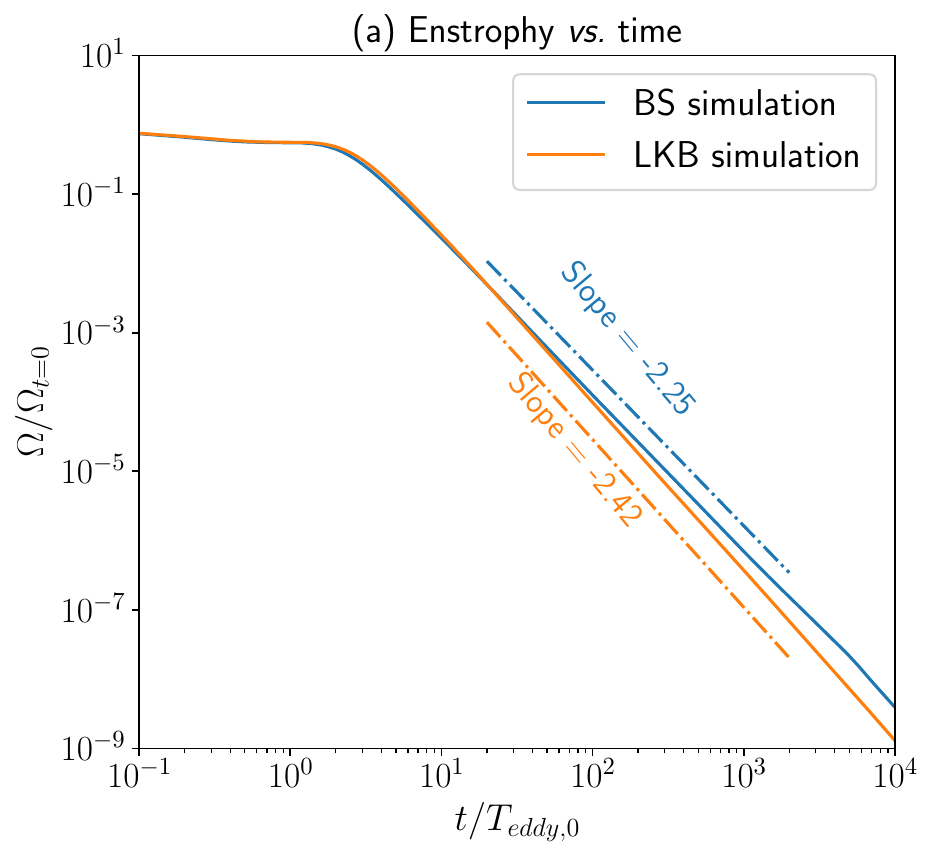}
    \end{subfigure}
    \hfill % Adds horizontal space between subfigures
    \begin{subfigure}[b]{0.5\textwidth} % Adjust width as needed
        \centering
        \includegraphics[width=\textwidth]{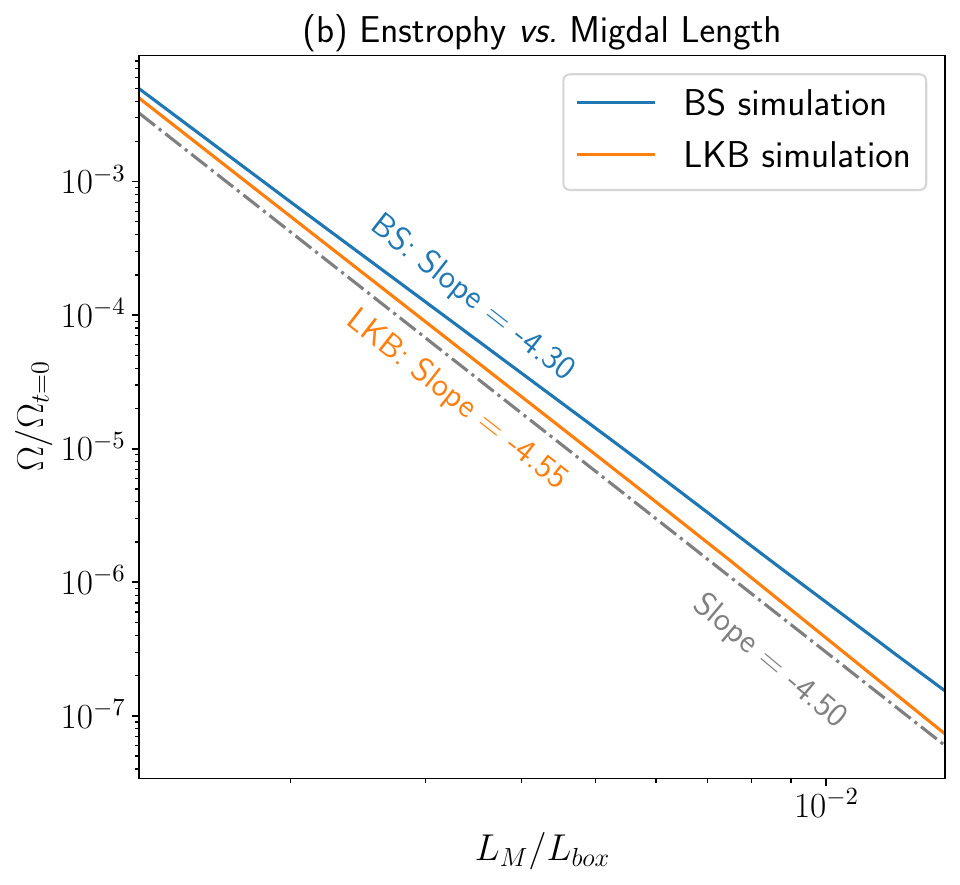}
    \end{subfigure}
    \caption{Evolution of enstrophy for the $Re_{\lambda}=145$ simulations. The solid blue lines represent the BS case, and the solid orange lines represent the LKB case. (a) Time evolution of normalized enstrophy $\Omega(t)/\Omega(0)$. The BS case decays with a slope of $-2.25$, consistent with $n=1.25$, while the LKB case decays with a slope of $-2.42$, consistent with $n \approx 10/7$. (b) Normalized enstrophy plotted against $L_M/L_{box}$. The dashed reference line indicates the theoretical slope of $-4.50$. The LKB case ($\approx -4.55$) aligns closer to this prediction than the BS case ($\approx -4.30$).}
    \label{fig:enstrophy}
\end{figure}

{Towards the end of \hyperref[sec:results]{\S3}, we have expressed the view that universality could be sought perhaps in terms of enstrophy, defined as $\Omega(t) = \int_0^{k_{\max}} k^2 E(k,t) dk$. We explore it here. To recapitulate, we had compared in \hyperref[subsec:migdal_prediction]{\S4b} the simulation results, for both BS and LKB initializations, with theory in terms of energy decay, length scale, and the slope of the second-order structure function. Here, we show in \hyperref[fig:enstrophy]{figure}~\ref{fig:enstrophy} the evolution of enstrophy against time (panel a) and against $L_M$ (panel b) for the $Re_{\lambda}=145$ simulations. (Needless to say, because of the high resolution employed, the enstrophy integral converges for all cases considered.)}

{As shown in \hyperref[fig:enstrophy]{figure}~\ref{fig:enstrophy}\hyperref[fig:enstrophy]{a}, the enstrophy decays as $\Omega(t) \sim t^{-2.25}$ for BS and $\sim t^{-2.42}$ for LKB. Since enstrophy is proportional to the energy dissipation rate $\epsilon$, it is expected to scale as $t^{-(n+1)}$, where $n$ is the energy decay exponent. The observed exponents (for some reasonable ranges of decay times) are consistent with this relation: for BS, $n \approx 1.25$ implies $\Omega \sim t^{-2.25}$, and for LKB, $n = 10/7$ implies $\Omega \sim t^{-2.43}$, which is very close to the measured $-2.42$.} The exponents differ slightly but measurably.

{\hyperref[fig:enstrophy]{Figure}~\ref{fig:enstrophy}\hyperref[fig:enstrophy]{b} plots the enstrophy directly against $L_M$. The data follow power laws of the form $\Omega \sim L_M^{\alpha}$, with $\alpha \approx -4.30$ for BS and $\alpha \approx -4.55$ for LKB.
Migdal's theory predicts an energy decay of $En \sim t^{-1.25}$ (implying $\Omega \sim t^{-2.25}$) and a length scale growth of $L_M \sim t^{0.5}$. Consequently, the theory predicts a scaling of $\Omega \sim L_M^{-4.5}$. Interestingly, the LKB simulation shows a closer agreement with this specific prediction ($\alpha \approx -4.55$) than the BS simulation ($\alpha \approx -4.30$). This appears to be a result of compensating factors in the LKB case: while the enstrophy decay is faster than the theoretical $-2.25$ and the length growth is faster than the theoretical $0.5$, their ratio aligns closely with the theoretical prediction of $-4.5$. By contrast, the BS simulation matches the theoretical time-decay of enstrophy exactly ($-2.25$) but exhibits a slightly faster growth of the length scale ($0.53$), leading to a somewhat larger deviation in the $\Omega$ versus $L_M$ scaling.}

\section{Sensitivity of bulk parameters to low-Wavenumber truncation}
\label{sec:5}
\subsection{Definitions of bulk parameters}{Discussions so far have shown that the flows initialized\enlargethispage{-12pt} with BS and LKB conditions exhibit distinct decay laws. This difference appears primarily owing to the respective $k^2$ and $k^4$ scalings in the low-wavenumber range. A decay law that depends strongly on such low wavenumbers may not have strong physical significance because, in practice, no physical system can be controlled to great accuracy at such low wavenumbers, causing the persistence of `boundary effects'. The question naturally arises: what if we filtered out the first few wavenumbers in the results on energy spectrum? Clearly, the effect of the initial low wavenumbers does not reside entirely in the same low wavenumbers at all subsequent times, but one may expect from the notion of the `permanence of large eddies' that a major part of initially low wavenumbers resides in low wavenumbers subsequently, as well.}

{We now define the `bulk' energy ($E_b$), bulk enstrophy ($\varOmega_b$) and the bulk version of the Migdal length ($L_b$) by integrating the spectrum from a lower-bound cutoff $k_0$ to the maximum resolved wavenumber $k_{\max}$:}
\begin{align}
    E_b(t) &= \int_{k_0}^{k_{\max}}E(k,t)dk\\
    \varOmega_b(t) &= \int_{k_0}^{k_{\max}}k^2 E(k,t)dk\\
    L_b(t) &= \frac{\int_{k_0}^{k_{\max}}k E(k,t)dk}{\int_{k_0}^{k_{\max}}k^2 E(k,t)dk}
\end{align}
{where $k_0 \in \{0, 1, \dots, 7\}$. This section details the sensitivity of these bulk parameters and their resulting power-law evolution to the choice of the cutoff $k_0$.} (We have used the notation `bulk' mostly in recognition of the fact that the definitions exclude `boundary' effects.)

\subsection{The influence of $k_0$ on decay evolution}
\begin{figure}[]
    \centering
    % --- Row 1 ---
    \begin{subfigure}[b]{0.48\textwidth}
        \centering
        \includegraphics[width=\linewidth]{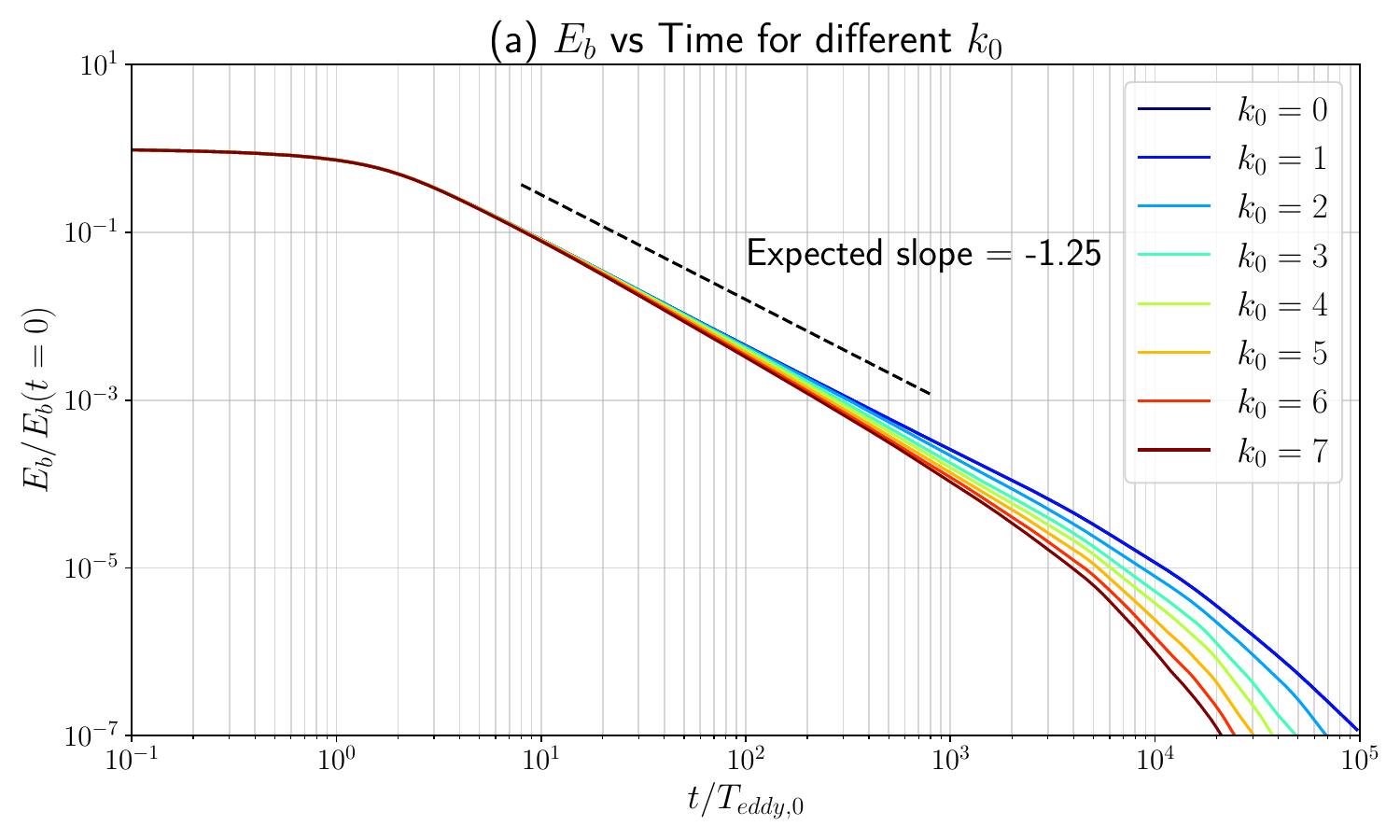}
        % \caption{Normalized bulk energy \textit{vs.} time for the BS case.}
        % \label{fig:En_k2}
    \end{subfigure}
    \hfill % Adds flexible space between images
    \begin{subfigure}[b]{0.48\textwidth}
        \centering
        \includegraphics[width=\linewidth]{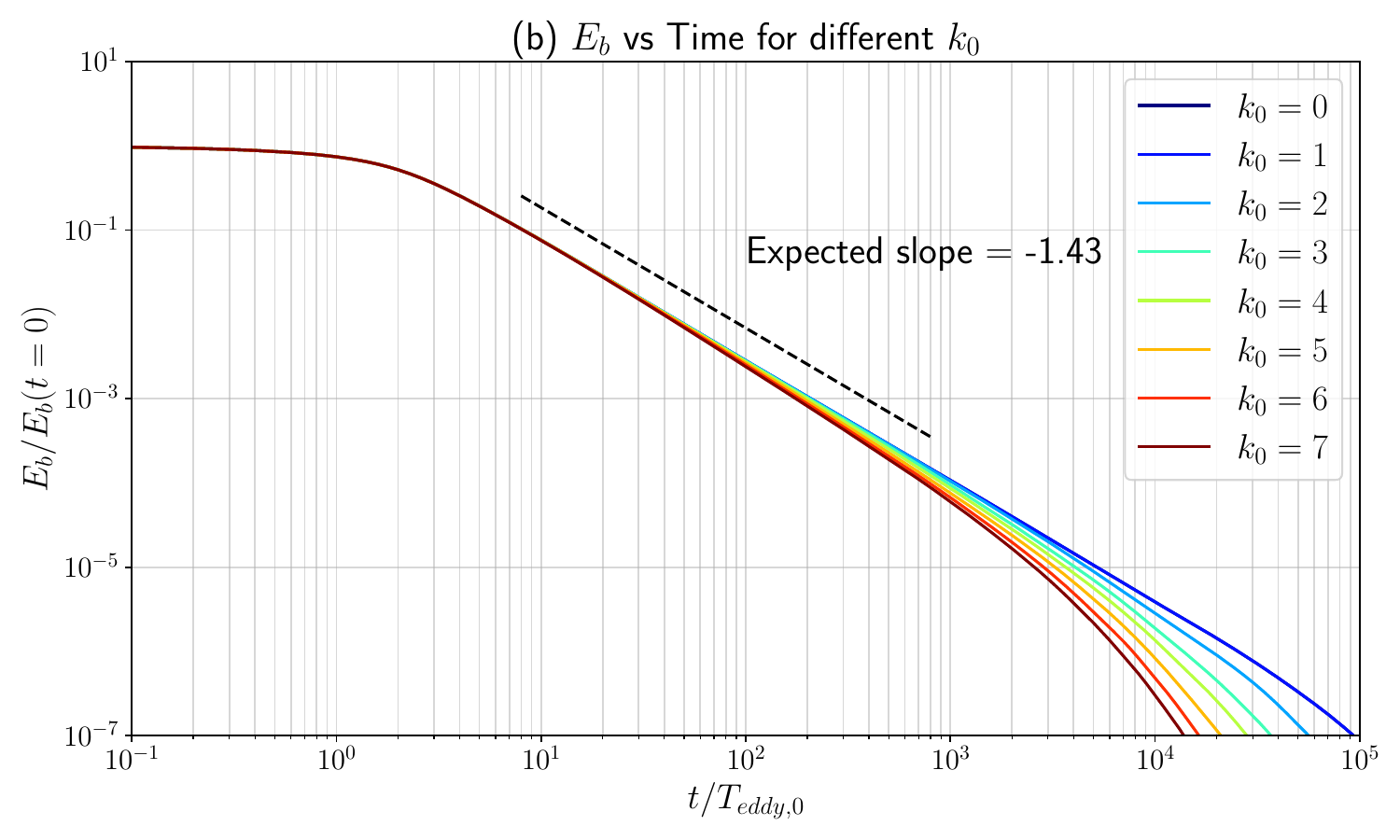}
        % \caption{Normalized bulk energy \textit{vs.} time for the LKB case.}
        % \label{fig:En_k4}
    \end{subfigure}
    
    \par\bigskip % Forces a new row and adds vertical space
    
    % --- Row 2 ---
    \begin{subfigure}[b]{0.48\textwidth}
        \centering
        \includegraphics[width=\linewidth]{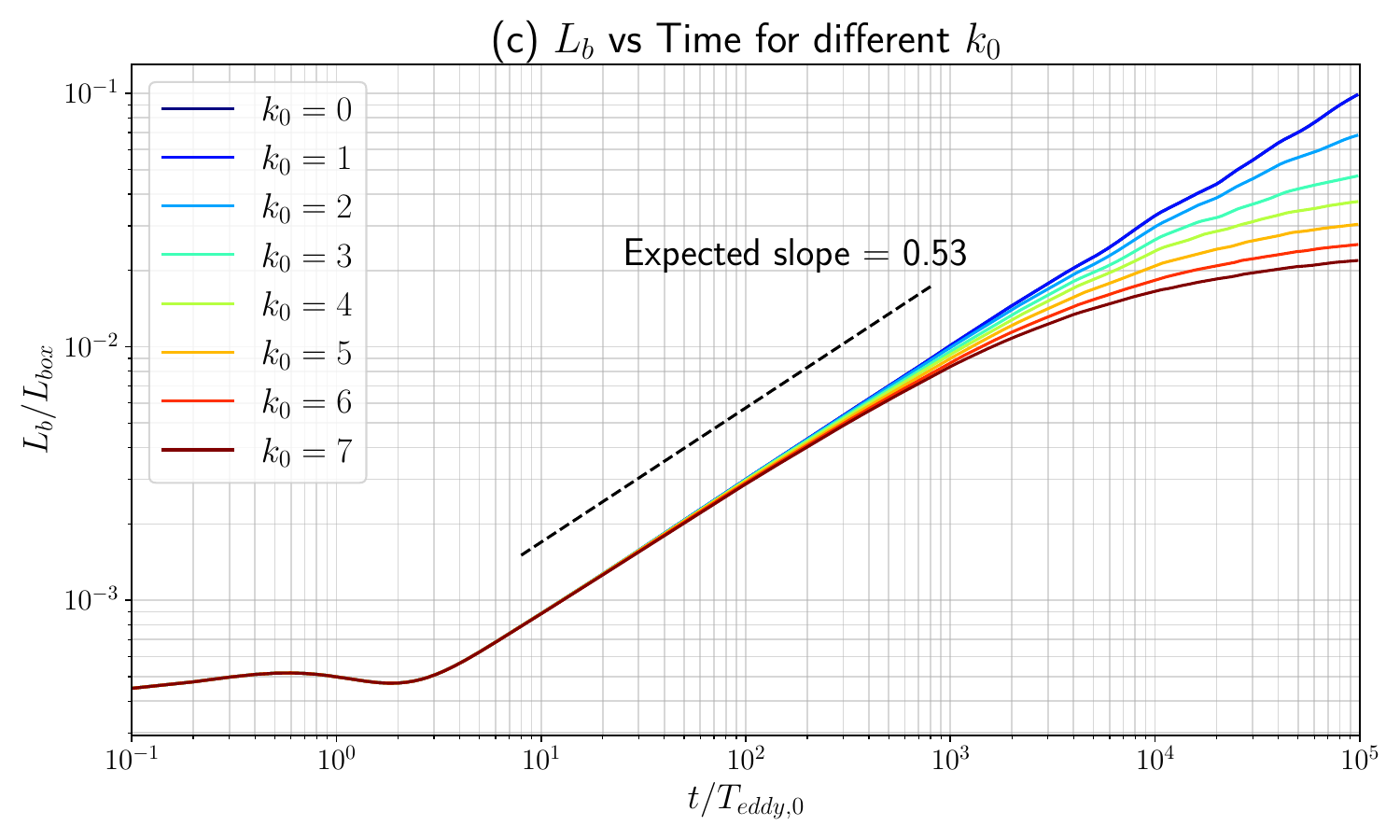}
        % \caption{Migdal length for BS case.}
        % \label{fig:fig3}
    \end{subfigure}
    \hfill
    \begin{subfigure}[b]{0.48\textwidth}
        \centering
        \includegraphics[width=\linewidth]{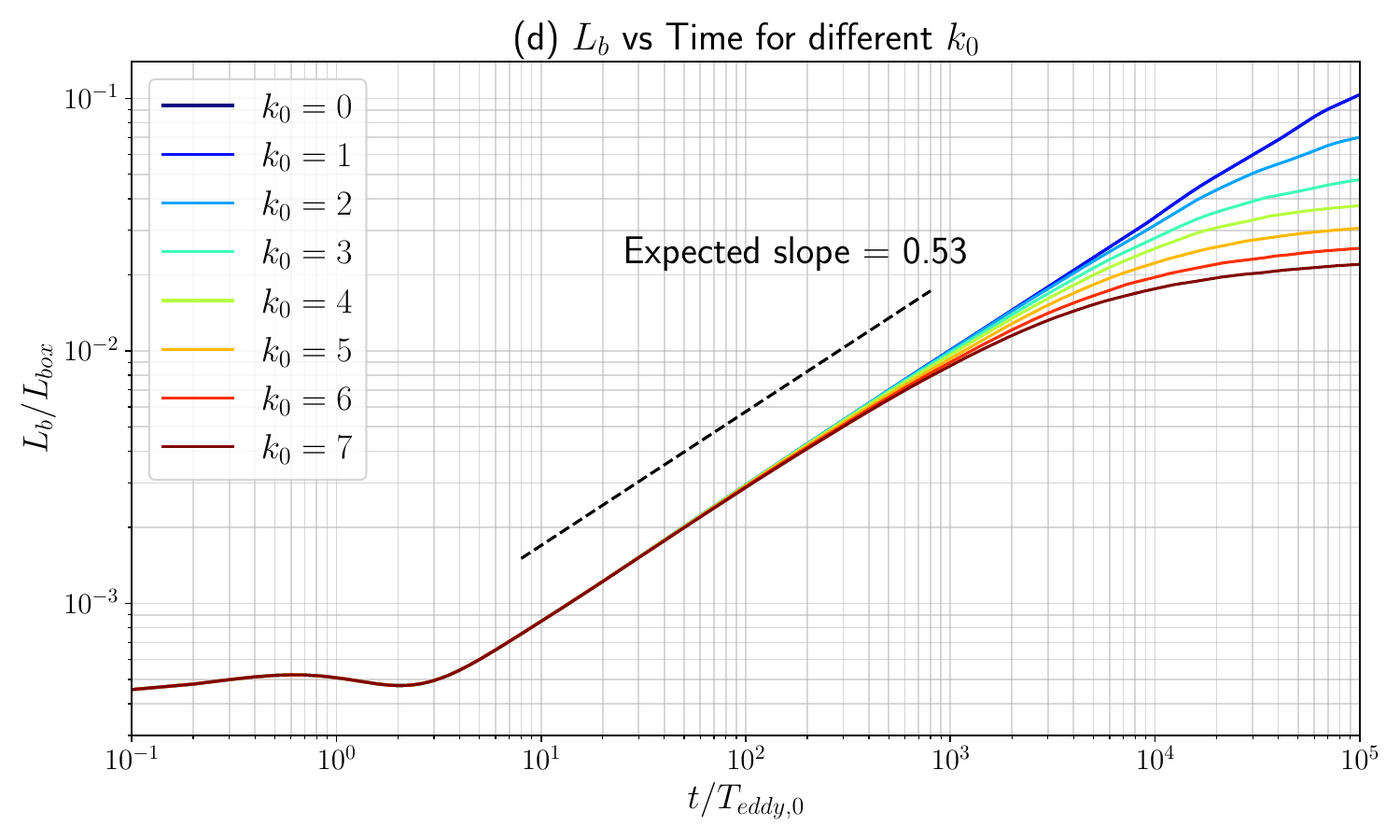}
        % \caption{Migdal length for LKB case.}
        % \label{fig:fig4}
    \end{subfigure}
    
    \par\bigskip % Forces a new row and adds vertical space

    % --- Row 3 ---
    \begin{subfigure}[b]{0.48\textwidth}
        \centering
        \includegraphics[width=\linewidth]{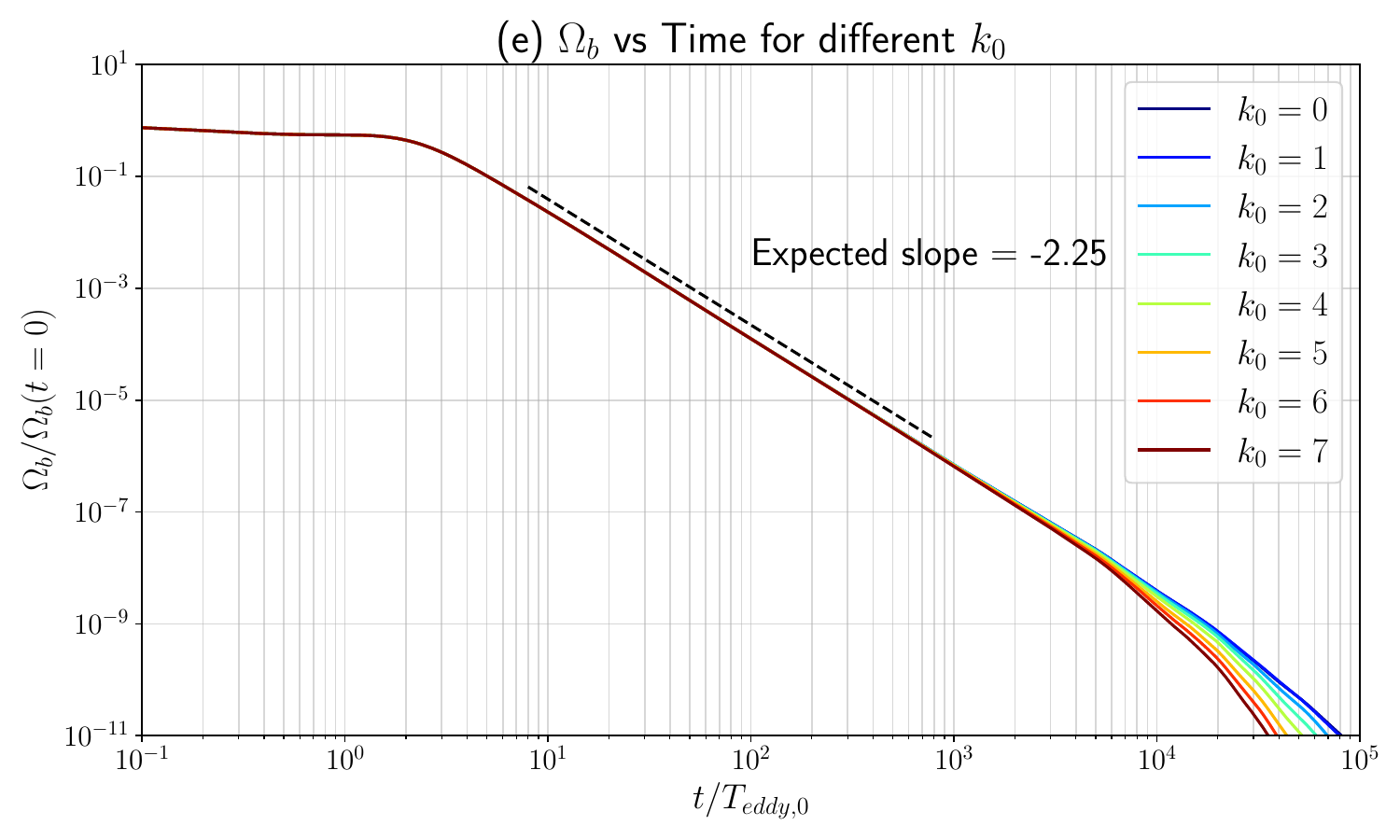}
        % \caption{Enstrophy for BS case.}
        % \label{fig:fig5}
    \end{subfigure}
    \hfill
    \begin{subfigure}[b]{0.48\textwidth}
        \centering
        \includegraphics[width=\linewidth]{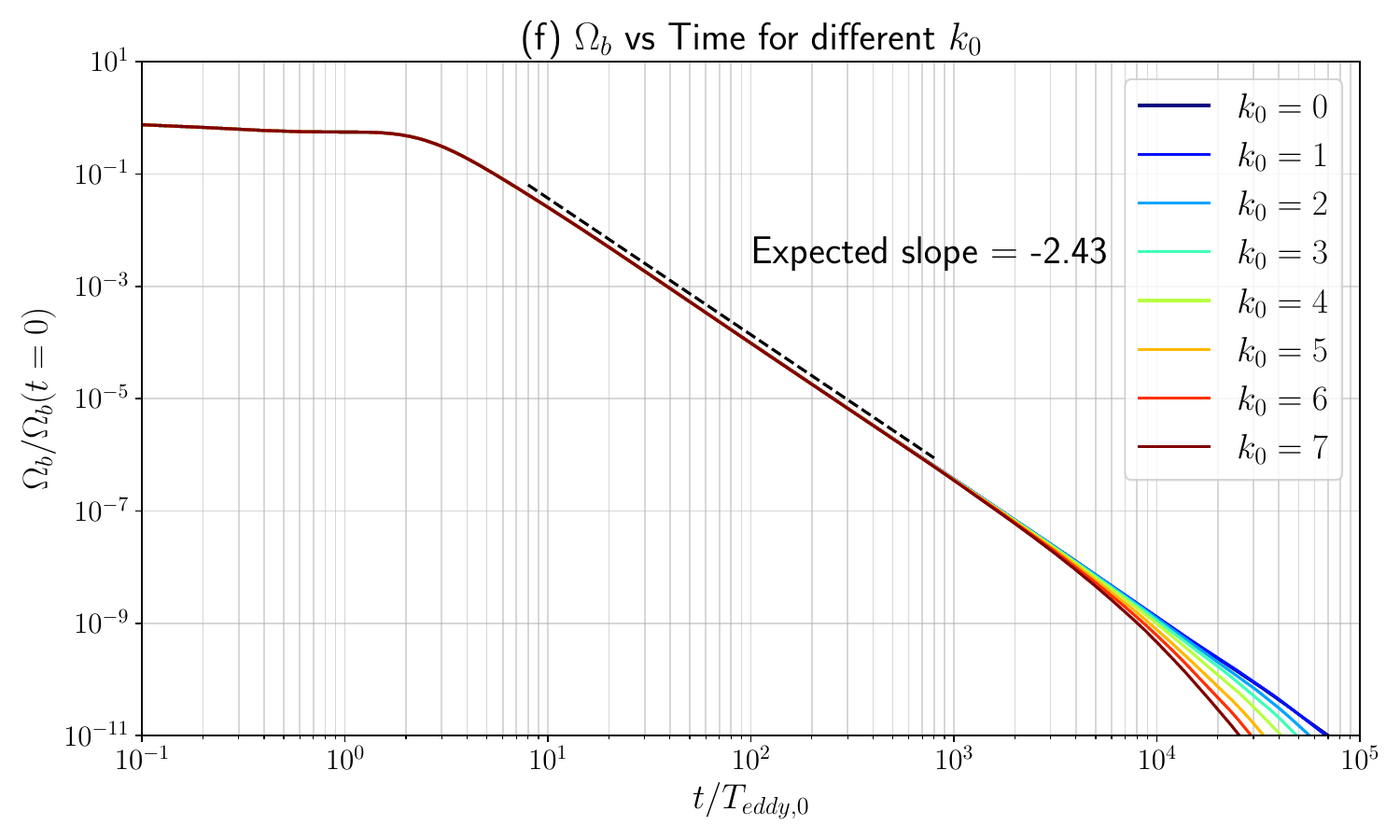}
        % \caption{Enstrophy for LKB case.}
        % \label{fig:fig6}
    \end{subfigure}
    
    \caption{Time evolution of bulk turbulence parameters for $Re_{\lambda}=145$, comparing simulations with BS initial conditions (left column) and LKB initial conditions (right column). Each plot displays curves for varying low-wavenumber cutoffs $k_0 \in \{0, 1, \dots, 7\}$. (a–b) are for the bulk energy ($E_b(t)$),  normalized by the initial value, for the BS case (a) and LKB case (b). As $k_0$ increases, the energy decay differs from the expected power law ({as shown by dashed line in the figures}) at progressively earlier times and shows a steeper decay rate. (c–d) are for $L_b(t)$:  BS (c) and LKB (d). Similar to the $E_b$, $L_b$ shows significant sensitivity to $k_0$, with the growth rate decreasing as larger number of low-wavenumber modes are excluded. (e–f) are for bulk enstrophy ($\Omega_b(t)$) for BS (e) and LKB (f). Compared to energy and length scales, enstrophy is relatively insensitive to the choice of $k_0$, maintaining a consistent power-law behavior for longer durations.}
    \label{fig:k0_effect}
\end{figure}

We examined the time evolution of $E_b(t)$, $\Omega_b(t)$ and $L_b(t)$ for simulations at $Re_\lambda = 145$. \hyperref[fig:k0_effect]{Figure}~\ref{fig:k0_effect} illustrates these trends for BS (left column) and LKB (right column) spectra:
The first row (\hyperref[fig:k0_effect]{figure}~\ref{fig:k0_effect}\hyperref[fig:k0_effect]{a,b}) is for $E_b(t)$ versus $t$;
second row (\hyperref[fig:k0_effect]{figure}~\ref{fig:k0_effect}\hyperref[fig:k0_effect]{c,d}) for $L_b(t)$ versus $t$
and third row (\hyperref[fig:k0_effect]{figure}~\ref{fig:k0_effect}\hyperref[fig:k0_effect]{e,f}) for $\Omega_b(t) $ versus $t$.

Not surprisingly, $E_b(t)$ and $L_b(t)$ are more sensitive to increases in $k_0$ than $\Omega_b(t)$. As $k_0$ increases, $E_b$ and $L_M$ depart from their respective unfiltered power laws at increasingly early stages of the decay---obviously because the fraction of energy contained within the low-wavenumber modes $(k < k_0)$ grows over time. Understandably, $\Omega_b$ remains relatively stable. One consequence is that the exponents for the composite relations---$\Omega_b$ versus $L_M$ and $E_b$ versus $L_M$---increase with $k_0$ and cause the measured exponents to depart further from the theoretical values of 4.5 and 2.25, respectively.

We used least-square fits to obtain power-law slopes within the interval $10 < t/T < 1000$. (Our main results utilized a window of $20 < t/T < 2000$, but we shifted it slightly for this analysis.) \hyperref[tab:BS_10-1000]{Tables}~\ref{tab:BS_10-1000} and~\ref{tab:LKB_10-1000} provide the new exponents for BS and LKB simulations for different $k_0$.

\begin{table}[]
\centering
\begin{tabular}{|c|c|c|c|c|c|}
\hline
$k_0$ & $\frac{d \log E_b}{d \log t}\times 10^{0}$ & $\frac{d \log \Omega_b}{d \log t}\times 10^{0}$ & $\frac{d \log L_M}{d \log t}\times 10^{-1}$ & $\frac{d \log \Omega_b}{d \log L_M}\times 10^{0}$ & $\frac{d \log E_b}{d \log L_M}\times 10^{0}$\\ \hline
0     & -1.246 & -2.261 & 5.285 & -4.278 & -2.357   \\ \hline
1     & -1.246 & -2.261 & 5.285 & -4.278 & -2.357  \\ \hline
2     & -1.299 & -2.262 & 5.227 & -4.327 & -2.485  \\ \hline
3     & -1.359 & -2.263 & 5.114 & -4.425 & -2.655  \\ \hline
4     & -1.390 & -2.266 & 5.033 & -4.500  & -2.760  \\ \hline
5     & -1.418 & -2.270 & 4.936 & -4.595  & -2.869  \\ \hline
6     & -1.442 & -2.274 & 4.838 & -4.695  & -2.975  \\ \hline
7     & -1.463 & -2.279 & 4.751 & -4.790  & -3.073  \\ \hline
\end{tabular}
\caption{Slopes of different curves for different values of $k_0$ for the BS simulation ($Re_{\lambda} = 145$).}
\label{tab:BS_10-1000}
\end{table}

\begin{table}[]
\centering
\begin{tabular}{|c|c|c|c|c|c|}
\hline
$k_0$ & $\frac{d \log E_b}{d \log t}\times 10^{0}$ & $\frac{d \log \Omega_b}{d \log t}\times 10^{0}$ & $\frac{d \log L_M}{d \log t}\times 10^{-1}$ & $\frac{d \log \Omega_b}{d \log L_M}\times 10^{0}$ & $\frac{d \log E_b}{d \log L_M}\times 10^{0}$ \\ \hline
0     & -1.421 & -2.422 & 5.368 & -4.511 & -2.646 \\ \hline
1     & -1.421 & -2.422 & 5.368 & -4.511 & -2.646 \\ \hline
2     & -1.428 & -2.422 & 5.361 & -4.517 & -2.664 \\ \hline
3     & -1.459 & -2.423 & 5.301 & -4.563 & -2.747 \\ \hline
4     & -1.491 & -2.425 & 5.235 & -4.631 & -2.846 \\ \hline
5     & -1.527 & -2.429 & 5.135 & -4.727 & -2.970 \\ \hline
6     & -1.554 & -2.433 & 5.041 & -4.822 & -3.078 \\ \hline
7     & -1.578 & -2.438 & 4.954 & -4.915 & -3.179 \\ \hline
\end{tabular}
\caption{Slopes of different curves for different values of $k_0$ for the LKB simulation ($Re_{\lambda} = 145$).}
\label{tab:LKB_10-1000}
\end{table}

\section{Concluding remarks}
\label{sec:conclusion}
In this work, we have revisited the classical problem of decaying homogeneous isotropic turbulence using DNS for an unprecedented length of time as well as grid resolution. By employing a dynamic grid modification strategy, we extended (in some cases) the simulation time to over $2 \times 10^5$ initial eddy turnover times while maintaining high spectral resolution ($k_{\max}\eta \ge 3$). We surpassed the limitations of previous studies---insufficient duration and finite-domain effects---and isolated the asymptotic decay regimes. We also utilized these data particularly to evaluate the predictions of the recent loop-space theory of turbulence proposed by Migdal \cite{Migdal_theory}, which posits a universal decay governed by the Euler ensemble. Our analysis reveals a nuanced picture of the universality of energy decay because of how the initial large structure of the flow shapes the decay---as we outline here.

For the BS regime ($E(k) \sim k^2$ for small $k$), the simulation results for energy decay and the length scale align closely with Migdal's theoretical predictions. This is particularly true for his new predictions for the alternative length scale $L_M$ and the shape of the local slope of the structure function, $\zeta_2(r,t)$. However, for the LKB regime ($E(k) \sim k^4$ for small $k$), we observed a robust decay exponent of $n \approx 10/7$, which persists for thousands of turnover times, as expected from the classical LKB theory.  The Euler ensemble solution does not take account of this instance.

In our initial DNS study of some 4 years ago \cite{john-P-john}, we were struck by the level of non-universality of energy decay. This study is more complete and confirms that there are inherent complexities to seeking a universal decay exponent for the energy. The energy decay behaves as expected for classical results as long as the low wavenumber behaviour is controlled, but the situation is unpredictable if neither the $k^2$ nor the $k^4$ behaviour is initialized. (Whether the notion of permanence of large eddies precludes more general behaviours in an experiment is unclear.)

Given this situation, how should one frame the success of Migdal's theory? On the face of it, the simulation results demonstrate the limitations to the theory's universality for energy decay. It is possible that the observed decay exponent in an experiment can be regarded as $-5/4$ plus a correction term that emanates from `boundary effects' (as discussed in \hyperref[sec:5b]{\S5b}). Though details of this theoretical structure have not yet been developed fully, its elements are already present in Migdal's theory \cite{Migdal_theory}.

This feature aside, the internal structure of the flow---manifested in the growth of the Migdal length and the spectral slope $\zeta_2$---agrees with theoretical predictions. Together, these findings suggest that, while the global decay rate is sensitive to large-scale behaviour, universality may yet prevail for the internal relations {and to the decay of high wavenumber properties such as enstrophy}. The lack of universality in large scale properties such as energy will not surprise experts on critical phenomena (for example). If this is the correct view, it is possible that the question of universal decay exponents may have been poorly posed all along. We have explored the results for enstrophy decay, and also considered the consequence of excluding the lowest wavenumbers in the computation of energy, enstrophy and length scales. The Euler ensemble captures these features better, and can thus be said to lay an important foundation stone for a complete theory.

How should one interpret the close correspondence with the classical BS result on decay (which is conditioned on certain initial conditions) and Migdal's theory that purports to be universal? We merely note that the $k^2$ spectrum for the BS case corresponds to an equipartition of energy at low wavenumbers, and hence may be more general than initially conceived.

Future theoretical work needs to better reconcile how the universal internal structure (which appears robust for both regimes) coexists with distinctly different energy decay rates. On the simulations side, the need still exists for a small number of targeted and well-controlled simulations at higher Reynolds numbers.

\textit{Note added in proof}: In the six months that have elapsed since the writing of this paper, Migdal has developed his theory further, so it is now possible to give a summary that is more succinct. Further comparisons with the theory will be discussed elsewhere.

\vspace*{18pt}
\dataccess{The data  used in the paper, as well as the Python scripts required to reproduce all figures in the manuscript, including a README with usage instructions, can be found at \cite{rodhiya}.}

\disclaimer{We have not used AI-assisted technologies in creating this article.}

\noindent\textbf{Authors’ contributions.} A.R.: simulations, writing the original draft, and revising; K.R.S.: conceptualization, methodology, resources, supervision, writing the original draft and revising.

Both authors gave final approval for publication and agreed to be held accountable for the work performed therein.

\begin{conflict}
We declare we have no competing interests.
\end{conflict}

\funding{No funding has been received for this article. }

\ack{We owe our sincere thanks to Alexander Migdal for many patient discussions on his theoretical work, and for helpful comments on an initial draft. We are grateful to Diego Donzis, Shilpa Sajeev and John P. John for their generous insights while these simulations were begun, and thank Gregory Eyink, Snezhana Abarzhi and Sachin Bharadwaj for useful discussions. We acknowledge our debt to many previous workers on the topic, though we have not cited all their work. Some of the simulations were done on the cluster at New York University, which we acknowledge. Some simulations were also performed using the computational resources at the Texas Advanced Computing Center through ACCESS allocation CTS110029. For the additional simulations  performed on Shaheen III, the Supercomputing Laboratory of the King\break Abdullah University of Science and Technology (KAUST) in Saudi Arabia, we thank David Keyes with pleasure for his scientific leadership.}

% \noindent\textbf{Theme.}{ One contribution of 19 to a theme issue `Frontiers of turbulence and statistical physics'.}

%%%%%%%%%% Insert bibliography here %%%%%%%%%%%%%%

% \bibliographystyle{RS} % Or another style like plain, unsrt, abbrv
% \bibliography{sample} % No .bib extension needed here

\end{document}